\documentclass[a4paper,11pt]{article}
\pdfoutput=1 % if your are submitting a pdflatex (i.e. if you have
\usepackage{jcappub} % for details on the use of the package, please
\usepackage[T1]{fontenc} % if needed
\usepackage[utf8]{inputenc}

\usepackage{amsmath,amssymb,amsfonts}
\usepackage{graphicx}
\usepackage{enumerate} % advanced enumerate environment
\usepackage{colordvi} % for color text
\usepackage{bm}% bold math
\usepackage{epsfig}
\usepackage{float}
\usepackage{verbatim}
\usepackage{subfig}
\usepackage{xcolor}
\usepackage{multirow,multicol}

\usepackage[colorinlistoftodos]{todonotes}
\usepackage{hyperref}
\newcommand{\bfv}{\mbox{\boldmath$v$}}
\newcommand{\bfx}{\mbox{\boldmath$x$}}
\newcommand{\bfk}{\mbox{\boldmath$k$}}

\newcommand{\bfr}{\mbox{\boldmath$r$}}

\newcommand{\B}{\color{black}}

\newcommand{\eqtext}{equation}
\newcommand{\eqstext}{equations}
\newcommand{\tabletext}{Table}
\newcommand{\figuretext}{Figure}
\newcommand{\sectext}{Sec.}

\title{Hybrid $P_{\ell}(k)$: general, unified, non-linear matter power spectrum in redshift space}
\author[a]{Benjamin Bose,\footnote{Corresponding author.}}
\author[b]{Hans A. Winther,}
\author[c,f]{Alkistis Pourtsidou,}
\author[d]{Santiago Casas,}
\author[a]{Lucas Lombriser,}
\author[e]{Qianli Xia,}
\author[e]{Matteo Cataneo}

\affiliation[a]{D\'epartement de Physique Th\'eorique, Universit\'e de Gen\`eve, 24 quai Ernest Ansermet, 1211 Geneva 4, Switzerland}
\affiliation[b]{Institute of Theoretical Astrophysics, University of Oslo, 0315 Oslo, Norway}
\affiliation[c]{School of Physics and Astronomy, Queen Mary University of London, Mile End Road, London E1 4NS, UK}
\affiliation[d]{AIM, CEA, CNRS, Université Paris-Saclay, Université Paris Diderot, Sorbonne Paris Cité, F-91191 Gif-sur-Yvette, France}
\affiliation[e]{Institute for Astronomy, University of Edinburgh, Royal Observatory, Blackford Hill, Edinburgh, EH9 3HJ, UK}
\affiliation[f]{Department of Physics and Astronomy, University of the Western Cape, Cape Town 7535, South Africa}

\emailAdd{benjamin.bose@unige.ch}
\emailAdd{h.a.winther@astro.uio.no}
\emailAdd{a.pourtsidou@qmul.ac.uk}
\emailAdd{santiago.casas@cea.fr}
\emailAdd{lucas.lombriser@unige.ch}
\emailAdd{qx211@roe.ac.uk}
\emailAdd{matteo@roe.ac.uk}

\abstract{Constraints on gravity and cosmology will greatly benefit from performing joint clustering and weak lensing analyses  on large-scale structure data sets. Utilising non-linear information coming from small physical scales can greatly enhance these constraints. At the heart of these analyses is the matter power spectrum. Here we employ a simple method, dubbed ``Hybrid $P_\ell(k)$'', based on the Gaussian Streaming Model (GSM), to calculate the quasi non-linear redshift space matter power spectrum multipoles. This employs a fully non-linear and theoretically general prescription for the matter power spectrum. We test this approach against comoving Lagrangian acceleration simulation measurements performed in GR, DGP and $f(R)$ gravity and find that our method performs comparably or better to the dark matter TNS redshift space power spectrum model {\B for dark matter. When comparing the redshift space multipoles for halos, we find that the Gaussian approximation of the GSM with a linear bias and a free stochastic term, $N$, is competitive to the TNS model.} Our approach offers many avenues for improvement in accuracy as well as further unification under the halo model.
}

\begin{document}
\maketitle
\flushbottom

\section{Introduction}
The next decade is set to be a very exciting one for observational cosmology. The standard model, $\Lambda$CDM, will come under scrutiny as analysis pipelines open up to the wealth of large-scale structure (LSS) data provided by Stage IV surveys. These surveys, such as DES\footnote{\url{https://www.darkenergysurvey.org/}} \cite{Abbott:2018xao},  DESI\footnote{\url{https://www.desi.lbl.gov/}} \cite{Aghamousa:2016zmz},  Euclid\footnote{\url{www.euclid-ec.org}}  \cite{Amendola:2016saw,Blanchard:2019oqi},  and LSST\footnote{\url{https://www.lsst.org/}} \cite{Chisari:2018vrw}, are already underway or due to commence in the next few years. This prompts the provision of accurate models for the LSS observables. These models do not only need to be \emph{accurate}; they also have to be \emph{quick} to compute for parameter estimation analyses to be feasible, and \emph{unbiased} with theoretical systematics well under control. Ideally, they also have to be general and make as few assumptions as possible, or in other words, they should be largely \emph{agnostic} about gravity and dark energy. Furthermore, while phenomenological models have their merits, an ideal model would provide a fully \emph{consistent} framework based on some \emph{fundamental} underlying assumptions, which can give predictions for a range of LSS observables, such as galaxy clustering in redshift space (see \cite{Hamilton:1997zq} for a review) and weak lensing (see \cite{Bartelmann:2016dvf} for a review). In this work, we take a step in this direction by proposing and testing a simple prescription that has potential to encompass the aforementioned qualities.
\\
\\
The main goal of LSS surveys is to constrain cosmological parameters within the $\Lambda$CDM model but also investigate exotic models and probe the nature of dark energy and gravity, which remain largely untested at the non-linear scales of LSS. The primary statistic that has been used so far in deriving such constraints is the 2-point correlation function of the cosmological matter distribution, or its Fourier analog, the power spectrum. A great amount of work has gone into modelling these statistics accurately deep into the non-linear regime of structure formation (see \cite{Bernardeau2002, Bernardeau2015} for reviews). The most encompassing and accurate method is using {\it N}-body simulations, and the most restricted is employing linear perturbation theory that is only valid at the largest physical separations. To overcome computational costs associated with running full {\it N}-body simulations, emulators for the power spectrum and correlation function have been constructed  \cite{Smith2003,Takahashi:2012em,Mead:2015yca,Mead:2016zqy,Smith2018,Knabenhans:2018cng,Giblin:2019iit, Winther:2019mus,Schneider:2019snl}. These are fantastic tools, being fast and highly accurate in their predictions. They do however hinge on the various assumptions of  the {\it N}-body simulations they are constructed upon, such as model of gravity or presence of (massive) neutrinos. 
\\
\\
While the matter power spectrum, $P_{\delta \delta}$, is at the core of many LSS observables,  additional modelling is required when comparing with data. In particular, for lensing we must integrate the matter power spectrum over the line of sight to get the convergence power spectrum, compressing information from a wide range of physical scales in doing so. This makes it essential to model the non-linear matter density correctly. On the other hand, for the clustering power spectrum one must make the transformation from real space to redshift space, since this is where actual observations are made. This transformation encodes the so-called redshift-space distortions (RSD), which induce anisotropies into the spectrum. Naturally, the transformation involves peculiar velocity information of the matter field. Combining lensing and RSD spectra in an optimal way is one of the primary goals of current and upcoming surveys. Having a unified framework in which one can specify a single prescription for the matter power spectrum for both observables is highly desirable. 
\\
\\
As already mentioned, the state-of-the-art modelling approach for the non-linear matter power spectrum when aiming to fit to data and perform parameter estimation analyses, is the $\Lambda$CDM emulator. In some recent works, a prescription for extending such $\Lambda$CDM specific, non-linear power spectra to beyond $\Lambda$CDM models has been proposed \cite{Cataneo:2018cic}. This method, based off the so-called `reaction' of non-linear spectra to modified non-linear dynamics, relies on the well studied halo model as well as perturbation theory (see \cite{Seljak:2000gq,Smith2003,Cooray2002} and \cite{Bernardeau2015} for reviews, respectively). These two frameworks are building upon some very fundamental assumptions that tie in well to the current cosmological paradigm.  
\\
\\
On the redshift space modelling side, a number of prescriptions for the clustering, redshift space power spectrum have been proposed over the past decade, many of which rely on perturbation theory \cite{Carlson:2009it,Baumann:2010tm,Carrasco:2012cv,Blas:2015qsi, Pietroni:2008jx} and are hence very restricted in the range of scales they can accurately model. The part-perturbative, part-phenomenological TNS model~\cite{Taruya:2010mx} has stood out from these as being flexible in terms of modelling gravity and dark energy, as being accurate when compared to {\it N}-body simulations within a sufficient range of scales for galaxy clustering \cite{Nishimichi:2011jm,Taruya:2013my,Ishikawa:2013aea,Zheng:2016zxc,Gil-Marin:2015sqa,Gil-Marin:2015nqa,Bose:2017myh,Bose:2016qun,Markovic:2019sva,Bose:2019ywu,Bose:2019psj,delaBella:2018fdb}, and optimal as it only makes use of one free nuisance parameter to model RSD. It has also been used in the recent BOSS survey analysis \cite{Beutler:2013yhm,Beutler:2016arn}. 
On the other hand, recent works have performed validation studies using simulations \cite{delaBella:2018fdb, Bose:2019psj, Bose:2019ywu} as well as a re-analysis of BOSS data \cite{Ivanov:2018gjr, DAmico:2019fhj} using the Effective Field Theory (EFT) of LSS \cite{Senatore:2014vja, Carrasco:2013mua, Lewandowski:2017kes}, which apart from including 1-loop perturbation calculations, introduces so-called ``UV counterterms", which help to capture better the broadband shape and the oscillations of the power spectrum at higher $k$ values. As a downside, this approach introduces several free parameters that have to be measured with data or simulations. Similarly, in \cite{Hand:2017ilm}, the authors propose a non-linear RSD model based on the halo model, but this comes with 13 additional parameters leading to worse marginalised cosmological constraints than a perturbation theory model like TNS. However these parameters are physically motivated, and hence such a model has potential for great improvement if priors can be placed on these modelling degrees of freedom and/or parameter degenerecies can be broken. Finally, the BOSS collaboration has recently used a modification of TNS to model the non-linearities in RSD and BAO \cite{Troster:2019ean, Sanchez:2016sas}. This model uses a different perturbation theory resummation model, which is Galilean invariant, has non-linear bias contributions to the bispectra and a slightly different, non-Gaussian Finger-of-God model \cite{Scoccimarro:2004tg}. Given the above discussion, one can impose as a minimal requirement that any proposed model for the redshift space power spectrum should at least meet the TNS model in terms of accuracy and flexibility.
\\
\\
In this work we concentrate on applying a prescription for the non-linear matter power spectrum, as used in lensing analyses, to calculate the redshift space power spectrum. While the TNS and other perturbative models for the RSD power spectrum are primarily concerned with modelling the non-linear transformation from real to redshift space, we apply a different approach based on  the so-called \emph{streaming models} for the RSD correlation function \cite{Fisher:1994ks,Reid:2011ar,Uhlemann:2015hqa}. The primary concern in the streaming model approach is providing an accurate model for the pairwise velocity probability distribution function and the non-linear matter correlation function. This gives a lot more flexibility in the modelling, and may offer a clearer road into the highly non-linear regime of structure formation. These two ingredients are  convolved to get the RSD correlation function. To get the RSD power spectrum, we will simply take the Fourier transform of the streaming model predictions. A key point of this `hybrid' $P_\ell(k)$ (HyPk) approach is that the ingredients of the streaming model - RSD and real space clustering - can be modelled independently in a more consistent fashion than perturbative methods. This makes it somewhat more theoretically consistent to incorporate non-linear real space matter power spectra predictions in the clustering component.
\\
\\
The paper is organised as follows. In \sectext~\ref{sec:theory} we briefly review the LSS observables of interest, we describe the Gaussian streaming model ingredients, and present our HyPk approach. In \sectext~\ref{sec:comparison} we compare the HyPk predictions
against COmoving Lagrangian Acceleration (COLA) measurements for dark matter. {\B Additionally, for $\Lambda$CDM, we extend tests of the HyPk approach against halo measurements from fully non-linear {\it N}-body simulations}. Finally, in \sectext~\ref{sec:conclusion} we summarise our results and conclude.

\section{Theory} \label{sec:theory}
We aim to provide a (quasi-) non-linear model prescription for the power spectrum in redshift space that relies on a fully non-linear matter power spectrum as input. This non-linear matter spectrum prescription can also be directly applied to compute lensing statistics. Having such a unified framework is one of the primary goals of upcoming surveys that aim to perform joint clustering and lensing analyses \citep{Euclid2019}. Beyond this, we also aim to keep our method as general as possible in terms of gravitational and dark energy modelling. We begin by presenting the observables of interest. 
\newline
\newline
\subsection{2-point redshift space and lensing statistics} 
Typically, models for the redshift space power spectrum follow a perturbative route \cite{Kaiser:1987qv,Scoccimarro:2004tg,Taruya:2010mx,GilMarin:2012nb,Senatore:2014vja,Ivanov:2018gjr}, thus greatly restricting the Fourier wave modes that these models can probe. This does however offer the benefit of being completely general in terms of gravity and dark energy. Furthermore, a perturbative treatment is one way to tackle the non-linear transformation from real to redshift space. 
\\
\\
The flagship model for the recent BOSS analysis of the redshift space clustering power spectrum \cite{Beutler:2016arn} is the so called TNS \cite{Taruya:2010mx} model, which was combined with the tracer bias model of McDonald and Roy \cite{McDonald:2009dh}. The model is given as 
 \begin{align}
 P^{\rm S}_{\rm TNS}(k,\mu) =& D_{\rm FoG}(\mu^2 k^2 \sigma_v^2)\Big[ P_{g,\delta \delta} (k,b_1,b_2,N) + 2 \mu^2 P_{g,\delta \theta}(k,b_1,b_2) +  \mu^4 P_{\theta \theta}^{\rm 1-loop} (k) \nonumber \\ & \qquad \qquad \qquad \qquad + b_1^3A(k,\mu) + b_1^4B(k,\mu) +b_1^2 C(k,\mu)  \Big], 
 \label{redshiftps}
 \end{align} 
where the superscript ${\rm S}$ denotes a redshift space quantity, $\mu$ is the cosine of the angle between $\bfk$ and the line of sight and $P_g$ are the 1-loop galaxy power spectra with the bias model of \cite{McDonald:2009dh} implicitly included; the $A$,$B$ and $C$ terms are perturbative RSD correction terms \cite{Taruya:2010mx}, while the prefactor, $D_{\rm FoG}$, is a phenomenological term for the Finger-Of-God effect. Here we choose a Lorentzian form based on the conclusions of \cite{Markovic:2019sva, Bose:2019psj}
\begin{equation}
    D_{\rm FoG}^{\rm Lor}(k^2\mu^2 \sigma_v^2) = \frac{1}{1 + (k^2\mu^2 \sigma_v^2)/2} \, ,
\end{equation}
where $\sigma_v$ is a free parameter and represents the velocity dispersion of the tracers. We refer the reader to \cite{Bose:2019psj} for the formulas for the perturbative components of the model, along with the explicit dependency on the model's free  parameters $\{\sigma_v, b_1,b_2,N\}$, where $b_1$ is the linear bias, $b_2$ is the second order bias term and $N$ is a stochasticity term. For the dark matter spectrum, we set $b_1 = 1, b_2 = N = 0$ which results in $P_{g,\delta \delta} = P_{\delta\delta}^{\rm 1-loop}$ and $P_{g,\delta \theta}=P_{\delta\theta}^{\rm 1-loop}$.
\newline
\newline
In particular, in real data analyses, the full anisotropic 2D-spectrum is usually decomposed into its multipole moments defined as
\begin{equation}
P^{\rm{S}}_\ell(k)=\frac{2\ell+1}{2}\int^1_{-1}d\mu P^{S}(k,\mu)\mathcal{P}_\ell(\mu),
\label{multipoles}
\end{equation}
where $\mathcal{P}_\ell(\mu)$ denote the Legendre polynomials. It is worth noting that in linear theory the monopole ($P_0$), quadrupole ($P_2$), and hexadecapole ($P_4$) are the only non-zero multipoles and hence contain all of the cosmological information. Non-linearities give rise to higher order multipoles. In the recent BOSS analysis \cite{Beutler:2016arn} $P_0$ and $P_2$ were considered up to the same scale $k_{\rm max}$ where the model can be safely used; $P_4$ was also considered but to a smaller $k_{\rm max}$. It has been found to add slightly more information at scales which one can safely apply the TNS model \cite{Bose:2019ywu}, and can help break degeneracies and improve constraints on cosmological parameters. Despite this, modelling of $P_4$ is known to be challenging and its inclusion can bias cosmological parameter estimates if not handled with care \cite{Beutler:2016arn,Markovic:2019sva}. We will omit $P_4$ in this work, but it will be included in future parameter estimation studies using the hybrid approach. 
\newline
\newline
On the other hand, the lensing convergence spectrum in general relativity (GR), under the Limber approximation \cite{1953ApJ...117..134L} is given by \cite{Kaiser:1996tp,Jain:1996st} 
\begin{equation}
C_{l}\left(z_{s}\right)=\frac{9 \Omega_{m}^{2} H_{0}^{4}}{4} \int_{0}^{z_{s}} d z \frac{g^{2}(z)(1+z)^{2}}{\chi^{2}(z) H(z)} P^{\rm NL}_{\delta \delta}(k=l / \chi(z), z) \, ,
\label{lensingconv}
\end{equation}
 where $P(k, z)$ is the matter power spectrum at redshift $z$ and $g(z)$ is the lensing kernel. For a single source plane at $z=z_s$ we have $g(z)=\chi(z)\left(\chi(z)-\chi\left(z_{s}\right)\right) / \chi\left(z_{s}\right)$ where $\chi(z)$ is the co-moving distance. In general, modifications to gravity will alter the matter power spectrum, for example through the growth function, as well as the Weyl potential equation (see e.g.~\cite{Leonard:2015hha,Ishak:2018his,Joyce:2016vqv} for details).  
Because the matter power spectrum enters inside the integral in \eqtext~\ref{lensingconv}, we require accurate predictions for $P^{\rm NL}_{\delta \delta}(k=\ell/\chi)$ over a large range of scales. Thus, the perturbative prediction is not sufficient. We can instead use emulators to model $P^{\rm NL}_{\delta \delta}$ as these are sufficiently accurate to $k\sim10 \, h/{\rm Mpc}$ \cite{Knabenhans:2018cng}. Such emulators have largely been restricted to $\Lambda$CDM cosmologies, with the exception of \cite{Winther2019} which provides an emulator for the chameleon screened Hu-Sawicki model of $f(R)$ gravity \cite{Hu:2007nk}. 
\\
\\
Recently, \cite{Cataneo:2018cic} has proposed a means of obtaining $P^{\rm NL}_{\delta \delta}$ for general theories of gravity and dark energy at percent level accuracy, in a very fast and efficient way. This is done through modelling the so called \emph{reaction}, $\mathcal{R}$, which gives the non-linear correction to a $\Lambda$CDM power spectrum under dark energy or gravity modifications. The reaction is based on a combination of the halo model (see \cite{Cooray:2002dia} for a review) and 1-loop standard perturbation theory (SPT). This then allows one to compute the observable \eqtext~\eqref{lensingconv} for modifications to $\Lambda$CDM which do not modify the Weyl potential appreciably with respect to our measurement errors. 
\\
\\
With this in mind, we move on to present our procedure for predicting the redshift space power spectrum multipoles.  

%%%%%%%%%%%%%%%%%%%%%%%%%%%%%%%%%%%%%%%%%%

\subsection{Hybrid model}
Our aim is to use a single prescription for $P_{\delta \delta}^{\rm NL}$ in a model for the redshift space power spectrum multipoles, \eqtext~\eqref{multipoles}. {\B An important aspect of modelling the multipoles using perturbative methods is the transformation  of the density field from real to redshift space, which includes non-linear velocity information. An alternative to modelling the transformation of the density perturbation, is the so called \emph{streaming model} approach (see for example \cite{Fisher:1994ks,Reid:2011ar,Uhlemann:2015hqa}). These are configuration space models for the redshift space correlation function that pack all non-linear velocity information into the
particle velocity probability distribution function (PDF). Modelling this PDF and real space correlation function are essential for accurate predictions using this method. }
\\
\\
The streaming model approach to model the redshift space correlation function is to convolve the real space correlation function with this PDF
\begin{equation}
1+\xi^S(r_\sigma, r_\pi) = \int [1+\xi_g(r)] \mathcal{P}(r_\sigma,r_\pi, v_{12}, \sigma_{12}^2, \bar{\mu}_3, \bar{\mu}_4, \dots ) dr \, , \label{eq:steamingmodel}
\end{equation}
where $r_\sigma$ and $r_\pi$ are the separations perpendicular and parallel to the line of sight and $\xi_g$ is the real space correlation function of the dark matter tracer we are considering. If we are interested in dark matter only, $\xi_g = \xi$, the real space dark matter correlation function; $v_{12}$ is the mean of the PDF ($\mathcal{P}$) (the mean pair-wise infall velocity between tracer pairs) and $\sigma_{12}$ is the standard deviation of the distribution (the velocity dispersion of the tracers); {\B $\bar{\mu}_i$ for $i\geq 3$ are the higher order moments of the PDF.} 
\\
\\
This idea comes directly from the probabilistic definition of the correlation function in redshift space (see \cite{Fisher:1994ks,Bose:2018wxz}). The core assumption now is centred on what form the PDF takes and how can one model the parameters that characterise it.
\\
\\
Using a streaming model, and considering the dark matter field, we propose the following simple procedure to obtain the redshift space power spectrum multipoles of \eqtext~\eqref{multipoles}: 
\begin{enumerate}
\item
Perform a Fourier transform on the (non-linear) real space matter power spectrum to obtain $\xi(r)$:
\begin{equation}
\xi(r) = \frac{1}{2 \pi^2} \int_0^\infty dk k^2 P_{\delta \delta}^{\rm NL}(k) j_0(k r) \, .
\label{firstft}
\end{equation}
\item
Use this quantity to compute the streaming model multipoles: 
\begin{equation}
\xi_\ell^{(S)}(s)= 4 \pi \int^1_{-1}d\mu \xi^{S}(s,\mu)\mathcal{P}_\ell(\mu) \, .
\label{xi_multipoles}
\end{equation}
\item
Perform a Fourier transform on the streaming model multipoles to get the power spectrum multipoles:
\begin{equation}
P_\ell(k) = 4 \pi \int_0^\infty ds s^2 \xi_\ell^{(S)}(s) j_\ell(k s) \, ,
\label{finalft}
\end{equation}
where $j_\ell$ is the $\ell^{\rm th}$ order Bessel function. We note that we make use of the \texttt{mcfit}\footnote{\url{https://github.com/eelregit/mcfit}} package, which provides a Python version of the \texttt{FFTLog} algorithm \cite{Hamilton:1999uv}. This gives us an accurate and fast means of computing \eqtext~\eqref{firstft} and \eqtext~\eqref{finalft}.
\end{enumerate}
For the PDF, the simplest form we can take is a Gaussian, with which we only have the first two moments to consider (see \eqtext~(25) of \cite{Reid:2011ar})
\begin{equation}
1+\xi^S(r_\sigma, r_\pi) = \int [1+\xi(r)] e^{-[r_\pi -y -\mu v_{12}(r)]^2/[2 \sigma_{12}^2(r,\mu)]} \frac{dy}{\sqrt{2\pi \sigma^2_{12}(r,\mu)}}\, , 
\label{gsm_integral}
\end{equation}
where $y = \sqrt{ r^2- r_{\sigma}^2}$ is the real space separation parallel to the line of sight and $\mu = \hat{\ell} \cdot \hat{r}$ is the cosine of the angle between the line of sight ($\hat{\ell}$ being a unit vector in this direction) and the pair separation. This is the so called Gaussian streaming model (GSM) \cite{Reid:2011ar}. It has been shown to be a good approximation and is able to reproduce data well over a large range of scales as well as being competitive with other models for the RSD correlation function \cite{White:2014naa}.
\\
\\
The ingredients of the GSM are: $P_{\delta \delta}^{\rm NL}(k)$, $v_{12}(r)$ and $\sigma_{12}^2(r,\mu)$. Note that so far we have also been completely agnostic about gravity and dark energy. For these three ingredients one can adopt a host of prescriptions:
\begin{enumerate}
\item
$P_{\delta \delta}^{\rm NL}(k)$: Emulators, halo model \cite{Schmidt:2008tn,Schmidt:2009yj,Lombriser:2013eza,Barreira:2014zza,Lombriser:2014dua,Mead:2016zqy,Lombriser:2016zfz}, reaction approach for modified gravity and dark energy \cite{Cataneo:2018cic} or fitting formulae \cite{Smith:2002dz,Takahashi:2012em,Zhao:2013dza}.
\item 
$v_{12}(r, \mu)$ : 1-loop SPT \cite{Reid:2011ar,Bose:2017dtl}, various theoretical non-linear prescriptions \cite{Juszkiewicz:1998xf,Caldwell:1999uf,Hellwing:2014nma} or halo model approach \cite{Falco:2012ud,Falco:2013zya,Sheth:2000fe,Sheth:1995is,Diaferio:1996de}. 
\item
$\sigma_{12}^2(r,\mu)$:  1-loop SPT \cite{Reid:2011ar,Bose:2017dtl} or halo model approach \cite{Falco:2012ud,Falco:2013zya,Lombriser:2012nn,Sheth:2000fe,Sheth:1995is,Diaferio:1996de}. 
\end{enumerate}
One can of course also take measurements from simulations for all these ingredients (see for example \cite{Hellwing:2014nma}).
\\
\\
Using the reaction approach of \cite{Cataneo:2018cic} we have a general prescription for the first ingredient, $P_{\delta \delta}^{\rm NL}$. The mean pair-wise infall velocity and velocity dispersion can be modelled in a number of ways as indicated. In this work we employ the simplest, completely general procedure, and consider the 1-loop SPT predictions for these. We discuss these three ingredients in more detail next.  

\subsubsection{Reaction ($\mathcal{R}$) and non-linear matter power spectrum ($P_{\delta \delta}^{\rm NL}$)}
The non-linear power spectrum for a given theory of gravity or dark energy can be predicted as \cite{Cataneo:2018cic,Giblin:2019iit}
\begin{equation}
    P_{\delta \delta}^{\rm NL, real}(k;z) = \mathcal{R}(k;z) P^{\rm pseudo,NL}_{\delta \delta}(k;z).
\end{equation}
$P^{\rm pseudo,NL}_{\delta \delta}(k;z)$ is the so called \emph{pseudo} power spectrum. This is a non-linear $\Lambda$CDM power spectrum (as given by halofit \cite{Smith2003,Takahashi:2012em} for example), but with its linear clustering at redshift $z$ being equivalent to that of the modified gravity or dark energy theory under consideration. In the halofit example, this would simply be halofit run using the modified linear power spectrum as input.
\\
\\
The reaction, $\mathcal{R}$, is given by
\begin{equation}
\mathcal{R}(k;z) = \frac{ [(1-\mathcal{E})e^{-\frac{k}{k_\star}} + \mathcal{E}] P_L^{\rm real}(k,z) + P_{\rm 1h}^{\rm real}(k,z)}{P_L^{\rm real}(k,z) + P_{\rm 1h}^{\rm pseudo}(k,z)} \, , \label{eq:reaction}
\end{equation}
where $P_L$ is the linear power spectrum, and $P_{\rm 1h}$ is the 1-halo term coming from the halo-model (see \cite{Cooray:2002dia} for example). The quantity $\mathcal{E}$ is given by $\mathcal{E} = P_{\rm 1h}^{\rm real}(0.01;z)/P_{\rm 1h}^{\rm pseudo}(0.01;z)$. The superscript `real' denotes that a quantity is computed in the theory under consideration and `pseudo' denotes a $\Lambda$CDM quantity with the linear power spectrum replaced by the modified one, as done with $P^{\rm pseudo,NL}_{\delta \delta}(k;z)$. Finally, SPT enters via $k_{\rm \star}$, which is given by 
\begin{equation}
    k_{\rm \star} = - \bar{k} \left(\ln \left[ 
    \frac{A(\bar{k};z)}{P_L(\bar{k};z)} - \mathcal{E} \right] - \ln\left[1-\mathcal{E}\right]\right)^{-1} \, ,
\end{equation}
where 
\begin{equation}
    A(k;z) = \frac{[P_{\rm SPT}^{\rm real}(k;z)+ P_{\rm 1h}^{\rm real}(k,z)]
    [P_{\rm L}(k;z)+ P_{\rm 1h}^{\rm pseudo}(k,z)]}
    {P_{\rm SPT}^{\rm pseudo}(k;z)+ P_{\rm 1h}^{\rm pseudo}(k,z)} -  P_{\rm 1h}^{\rm real}(k,z) \, ,
\end{equation}
with $\bar{k} = 0.06 \, h/{\rm Mpc}$. $P_{\rm SPT}$ is the 1-loop SPT prediction for the matter-matter power spectrum and we have dropped the superscript from $P_L$, where it is assumed that it is always computed within the theory of dark energy or modified gravity under consideration. 
\\
\\
Finally, we note that we use the {\B publicly available} code {\tt ReACT} to compute \eqtext~\eqref{eq:reaction} \cite{react}. All halo model quantities are computed assuming a Sheth-Tormen mass function \cite{Sheth:1999mn,Sheth:2001dp}, a power law virial concentration and a Navarro-Frenk-White (NFW) halo density profile \cite{Navarro:1996gj}. We refer the interested reader to \cite{Cataneo:2018cic} for more details on the computation and ingredients. For the pseudo non-linear spectrum, $P^{\rm pseudo,NL}_{\delta \delta}(k;z)$, we use halofit with the modified linear power spectrum as an input, as detailed above. 

\subsubsection{Pair-wise infall velocity ($v_{12}$) and velocity dispersion ($\sigma_{12}^2$)}
Here we give some background on the first two moments of the particle pairwise velocity PDF.  These are given by the following correlations (see \eqtext~(26) and \eqtext~(28) of \cite{Reid:2011ar} or \eqstext~(2.33-2.46) of \cite{Bose:2017dtl}): 
\begin{equation}
[1+ \xi(r)] v_{12}(r) \hat{\bfr} = {\langle [1+\delta(\bfx)] [1+\delta(\bfx+\bfr)] [\bfv(\bfx + \bfr) - \bfv(\bfx)] \rangle} \, ,
\end{equation}
\begin{equation}
[1+ \xi(r)] \sigma_{12}^2 (r,\mu) =   {\langle [1+\delta(\bfx)] [1+\delta(\bfx+\bfr)] [v^\ell(\bfx + \bfr) - v^\ell(\bfx)]^2 \rangle} \, ,
\end{equation} 
where $\delta$ is the over-density perturbation, $\bfv$ is the velocity perturbation, and $v^\ell$ is the velocity projected along the line of sight, $v^{\ell} = \bfv \cdot \hat{\ell}$, with $\hat{\ell}$ being a unit vector along the line of sight. These can be modelled perturbatively using the above correlations (see the appendix of \cite{Reid:2011ar} for GR at 1-loop level or  \cite{Bose:2017dtl} for general theories at 1-loop level). Note that the perturbative approach for these components is not technically consistent with adopting a fully non-linear prescription for $\xi(r)$, but we can argue that such inconsistencies can be interpreted as inaccuracies in the PDF adopted. It has been shown that the Gaussian PDF is lacking when it comes to the very small scales of LSS \cite{Bianchi:2014kba,Bianchi:2016qen,Kuruvilla:2017kev,Valogiannis:2019nfz}. Further, the perturbative approach allows us to be very general when it comes to modelling these components. 
\newline
\newline
{\bf Isotropic contribution to velocity dispersion:} The correlation of $\sigma_{12}^2$ includes a term that is independent of the angle between the pair separation and the line of sight, $\mu$, nor does it depend on the separation $r$ itself. This correlation is given as 
\begin{equation}
\sigma_{\rm iso}^2 = \langle \delta(\bfx) v^\ell(\bfx)^2 \rangle,
\end{equation}
and in the literature this constant contribution to $\sigma_{12}^2$ has often been promoted to a free parameter, $\sigma_{\rm iso}$ \cite{Satpathy:2016tct,Reid:2012sw,Bose:2017dtl}. We do the same in our comparisons over the next couple of sections. The 1-loop SPT prediction for this quantity will be denoted as $\sigma_{\rm iso,pt}$.

\subsubsection{A note on bias}\label{sec:bias}
Since we are interested in using the fully non-linear power spectrum within the pipeline described by \eqtext~\eqref{firstft} to \eqtext~\eqref{gsm_integral}, tracer bias cannot be treated perturbatively in a consistent fashion. One approach would be to adopt a halo-model prescription for all three ingredients of the GSM model, and by doing so be able to include halo bias predictions. {\B For example, the Lagrangian bias using the peak background split formalism depends on the halo mass function \cite{Mo:1996cn,Sheth:1998xe,Mo:1995cs}, which is a key ingredient of the reaction in \eqtext~\eqref{eq:reaction}. This has been examined for modified gravity models in \cite{Valogiannis:2019xed} and implemented in a perturbative GSM modelling in \cite{Vlah:2016bcl,Valogiannis:2019nfz}.}  As this is an initial exploratory study and we are treating $\xi(r)$ non-linearly, such bias treatments are not considered here. Furthermore, the current version of our code is not optimised to fit multiple parameters efficiently. 
\\
\\
We still wish to provide some preliminary indication of the feasibility of HyPk in modelling the power spectra multipoles for biased tracers. For this purpose, we can attempt to include bias at the linear level in the comparisons ($g$ denotes a biased tracer) 
\begin{align}
    \xi_g \, \rightarrow& \,\, b_1^2 \xi \, , \\ 
    v_{12}^L \, \rightarrow& \,\, b_1 v_{12}^L \, , \\ 
    \sigma_{12}^L  \, \rightarrow& \,\, \sigma_{12}^L \, ,
\end{align}
where we only consider the linear theory predictions for $\sigma_{12}$ and $v_{12}$. Although not consistent, one can include linear bias at the 1-loop level for $v_{12}$ and $\sigma_{12}$ (see \cite{Reid:2011ar} for details). We experiment with this in the next section. 

%%%%%%%%%%%%%%%%%%%%%%%%%%%%%%%%%%%%%%%%%%%%%%%%%%%%%
\section{Comparisons to simulations}\label{sec:comparison}
{\B In this section we test the approach proposed here for the dark matter} redshift space multipoles against sets of COLA simulations \cite{Tassev:2013pn,Howlett:2015hfa, Valogiannis:2016ane,Winther:2017jof}. In particular, we use the \texttt{MG-PICOLA} code described in \cite{Winther:2017jof} to run simulations for three models of gravity: GR ($\Lambda$CDM), Hu-Sawicki $f(R)$ \cite{Hu:2007nk} gravity and the normal branch of DGP \cite{Dvali:2000hr} gravity. Each simulation has a cubic box of side $1024 \, {\rm Mpc}/h$ and $1024^3$ particles. The starting redshift is $z_{\rm ini}=49$ with initial conditions generated by second-order Lagrangian perturbation theory. Our initial power spectrum for all models of gravity use the following parameters: $\Omega_m = 0.281,  \Omega_b = 0.046, h=0.697, n_s=0.971$ and $\sigma_8(z=0)=0.844$. For $f(R)$ we take $|\bar{f}_{R0}| = 10^{-4}$ and for DGP we take $\Omega_{\rm rc} = 1/(2H_0r_c)^2 = 0.438$, which are both in the ruled out portion of parameter space (for example \cite{Lombriser:2009xg,Lombriser:2014dua,Barreira:2016ovx,Burrage:2017qrf,Cataneo:2014kaa}), but will give significant modifications to GR with which we can test the approach proposed here. %comment on screening because f4 has less prominent screening than f5 say.
%1412.5195 
\\
\\
For all simulations, we employ 10 independent realisations and then take the average spectra measurements for our comparisons. Further, for the redshift space measurements, each realisation measurement is the average over three line of sight directions. We measure the real space power spectrum, redshift space monopole and redshift space quadrupole at $z=0,0.5$ and $1$. Our error bars are chosen to reflect a Stage IV, Euclid-like survey \cite{Euclid2019}. We do this by using a linear theory covariance matrix between the multipoles (see the appendix of \cite{Taruya:2010mx} for example) for redshift space and a Fisher matrix projection \cite{Zhao:2013dza,Feldman:1993ky,Seo:2007ns} for the real space power spectrum. For these we assume a redshift bin volume of $V=4 \, {\rm Gpc}^3/h^3 $ and for the shot noise terms we assume a particle number density of $\bar{n}=1\times10^{-3}\, h^3/{\rm Mpc}^3$.  We note that linear theory covariance has been shown to reproduce {\it N}-body results up to $k\leq 0.3 \, h/\mbox{Mpc}$ at $z=1$ \cite{Taruya:2010mx}. A linear covariance also seems to work well at $z=0.5$ up to $k\leq 0.2 \, h/\mbox{Mpc}$ at $z=0.5$, as shown very recently in \cite{Sugiyama:2019ike}. We note that data analyses pipelines for Stage IV surveys will likely make use of both (validated) analytical approximations and numerical covariances constructed using mock catalogs.
\\
\\
\subsection{Real space spectra}
\begin{figure}[H]
\centering
  \includegraphics[width=\textwidth]{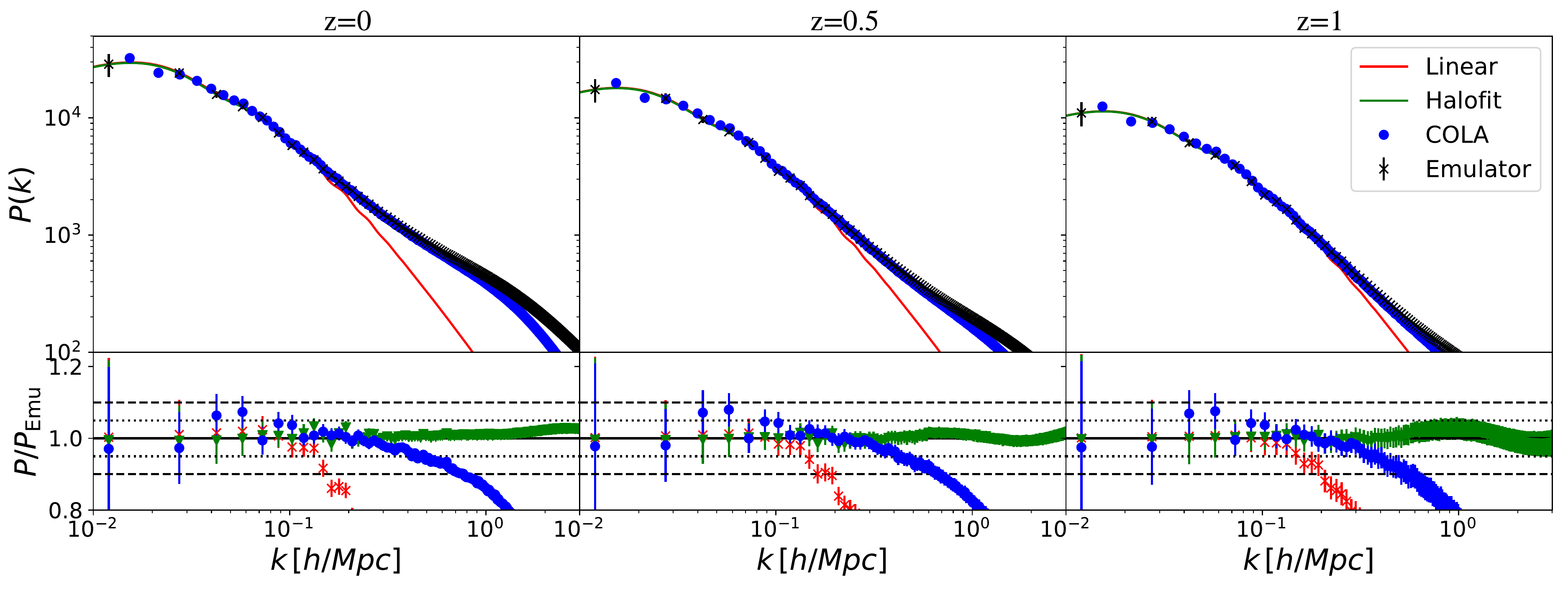}  
  \caption[]{{\bf Real space power spectrum in $\Lambda$CDM}: Top panels show the various prescriptions for the real space power spectrum, i.e. linear (red line), halofit (green line), Euclid emulator \cite{Knabenhans:2018cng} (black crosses) and the COLA measurement (blue circles). The bottom panels show the ratio of the top panel prescriptions to the emulator. The dotted and dashed lines indicate $5\%$ and $10\%$ deviation, respectively. The error bars are twice the Gaussian power spectrum variance which assumes stage IV survey like specifications as indicated in the main text.} 
\label{ps_gr}
\end{figure}

\begin{figure}[H]
\centering
  \includegraphics[width=\textwidth]{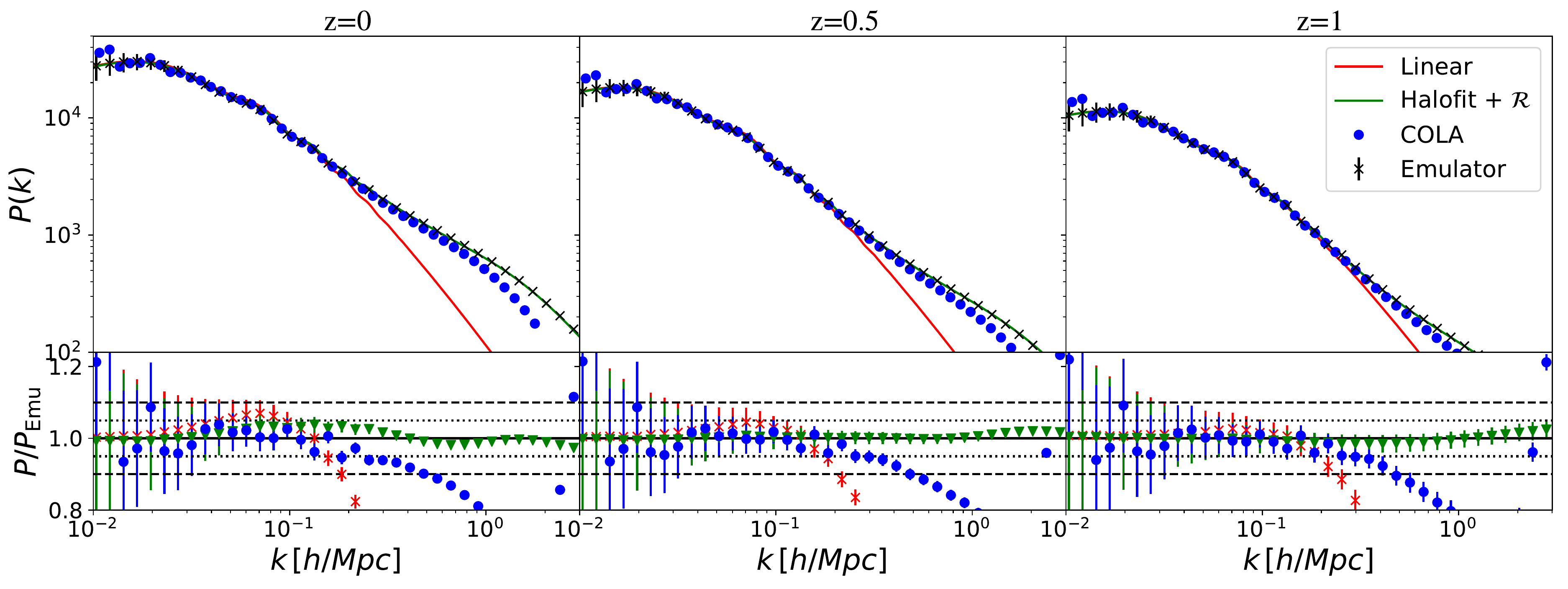}  
  \caption[]{{\bf Real space power spectrum in $f(R)$}: Same as \figuretext~\ref{ps_gr} but for $f(R)$. Here the halofit prescription is corrected by the halo model reaction $\mathcal{R}$ and the emulator is the one described in \cite{Winther2019}.}
\label{ps_fr}
\end{figure}

\begin{figure}[H]
\centering
  \includegraphics[width=\textwidth]{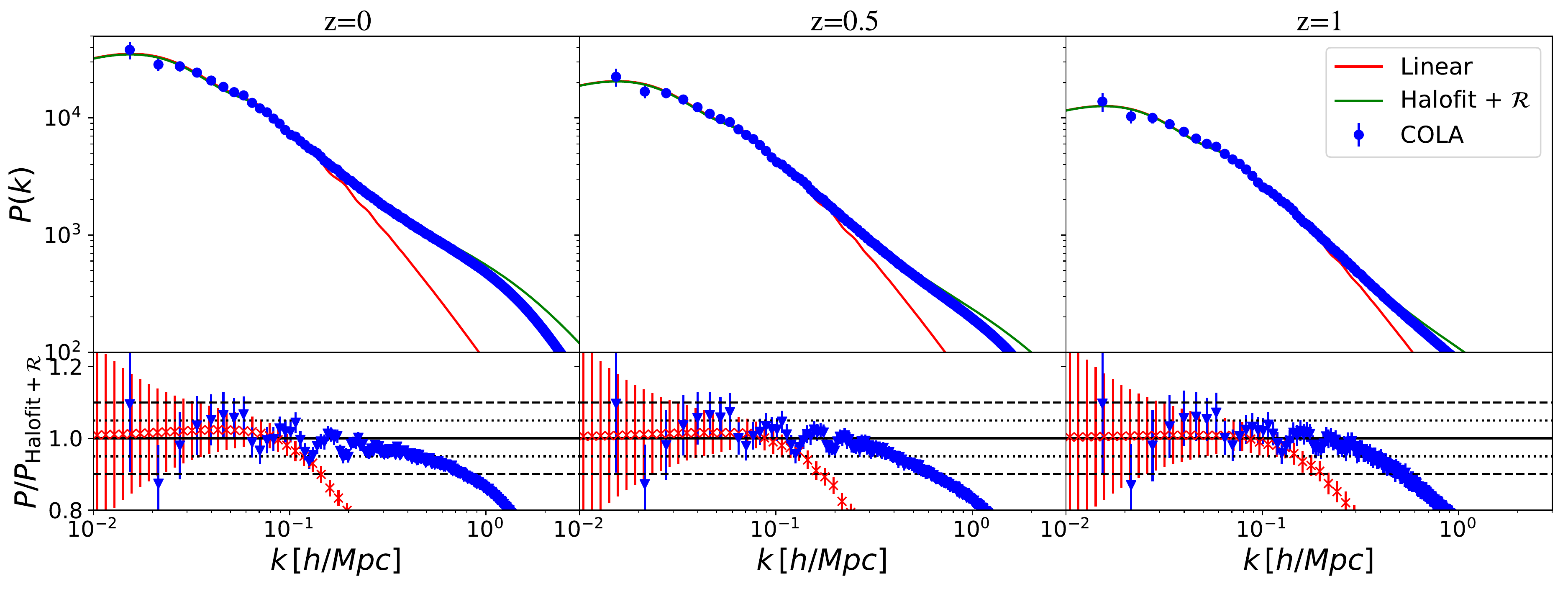}  
  \caption[]{{\bf Real space power spectrum in DGP}: Same as \figuretext~\ref{ps_fr} but for DGP.  The bottom panel shows the ratio of the COLA measurement and linear theory to the halofit prediction. Note again that the halofit prescription is corrected by the halo model reaction $\mathcal{R}$. }
\label{ps_dgp}
\end{figure}

We begin by comparing the non-linear real space dark matter spectra predictions: the emulator \cite{Knabenhans:2018cng,Winther2019}, the reaction \cite{Cataneo:2018cic} combined with halofit  \cite{Takahashi:2012em} for the pseudo spectrum, and COLA \cite{Winther:2017jof}. Among these, the emulators should be the most accurate and we use this as a reference (benchmark) spectrum. Similar comparisons have already been made in the literature (see the respective references of each prescription), but none comparing all three, particularly for modified gravity and in the non-linear regime ($k\leq3h/{\rm Mpc}$). These comparisons will be particularly useful in determining the gain in accuracy in the RSD multipoles by improving the accuracy of $P_{\delta \delta}^{\rm NL}$. 
\\
\\ 
\figuretext~\ref{ps_gr} shows comparisons of various predictions for the matter power spectrum at the three redshifts considered. The Euclid emulator \cite{Knabenhans:2018cng} is taken as our reference model for accuracy. We see that halofit offers sub $5\%$ accuracy for the scales considered, $k\leq 3 \, h/{\rm Mpc}$, while COLA is only accurate\footnote{Note that this accuracy is for the particular simulation setup used in this paper. The accuracy of power spectra generated using COLA simulations can be improved, but at the cost of increasing the runtimes for these simulations \cite{2016MNRAS.459.2327I}.} up to $k\sim0.5 \, h/{\rm Mpc}$. Similarly, \figuretext~\ref{ps_fr} shows the same results in $f(R)$ gravity, where we use the emulator of \cite{Winther2019} as our accuracy benchmark. Again, the halofit with a correction coming from the reaction prescription \cite{Cataneo:2018cic} offers sub $5\%$ accuracy with the emulator while COLA is only accurate to $k\sim0.5 \, h/{\rm Mpc}$. Finally, Figure~\ref{ps_dgp} shows the results for DGP. Here we do not have a benchmark emulator with which to compare the other prescriptions but we observe a deviation of COLA and the reaction-corrected-halofit prescription at similar scales at which COLA becomes inaccurate for GR and $f(R)$. 
\\
\\
It should be noted that an improvement in the pseudo power spectrum (here we use halofit) will offer improved accuracy in the modified gravity spectra \cite{Cataneo:2018cic,Giblin:2019iit}.  

%%%%%%%%%%%%%%%%%%%%%%%%%%%%%

\subsection{Redshift space spectra: dark matter}\label{sec:rsddm}
\begin{figure}[H]
\centering
  \includegraphics[width=\textwidth]{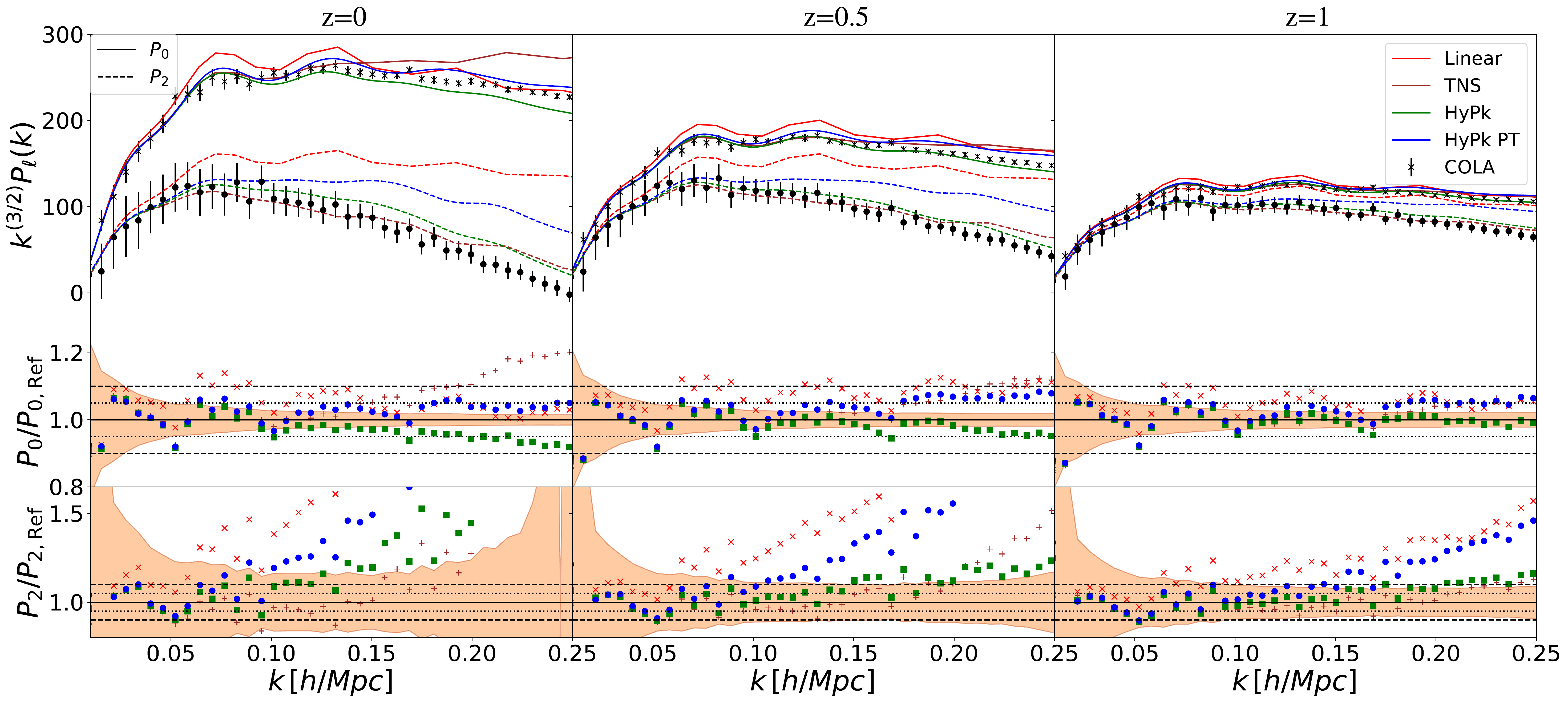}  
  \caption[]{{\bf Redshift space multipoles in $\Lambda$CDM}: Top panels show the COLA monopole (black crosses) and quadrupole (black circles) against the theoretical predictions. We show the theoretical predictions for the monopole (solid) and quadrupole (dashed) spectra using Kaiser (red), TNS (brown), HyPk with halofit with (green) and without (blue) $\sigma_{\rm iso}$ kept free. Middle and  bottom panels show the ratio of the theoretical prescriptions to the COLA measurements for $P_0$ and $P_2$ respectively, with the dotted and dashed lines indicating $5\%$ and $10\%$ deviation, respectively. The error bars and bands are computed using twice the diagonal components of the linear theory covariance matrix used in \eqtext~\ref{covarianceeqn} ($2\sigma$-errors).}
\label{p02_grb}
\end{figure}

\begin{figure}[H]
\centering
  \includegraphics[width=\textwidth]{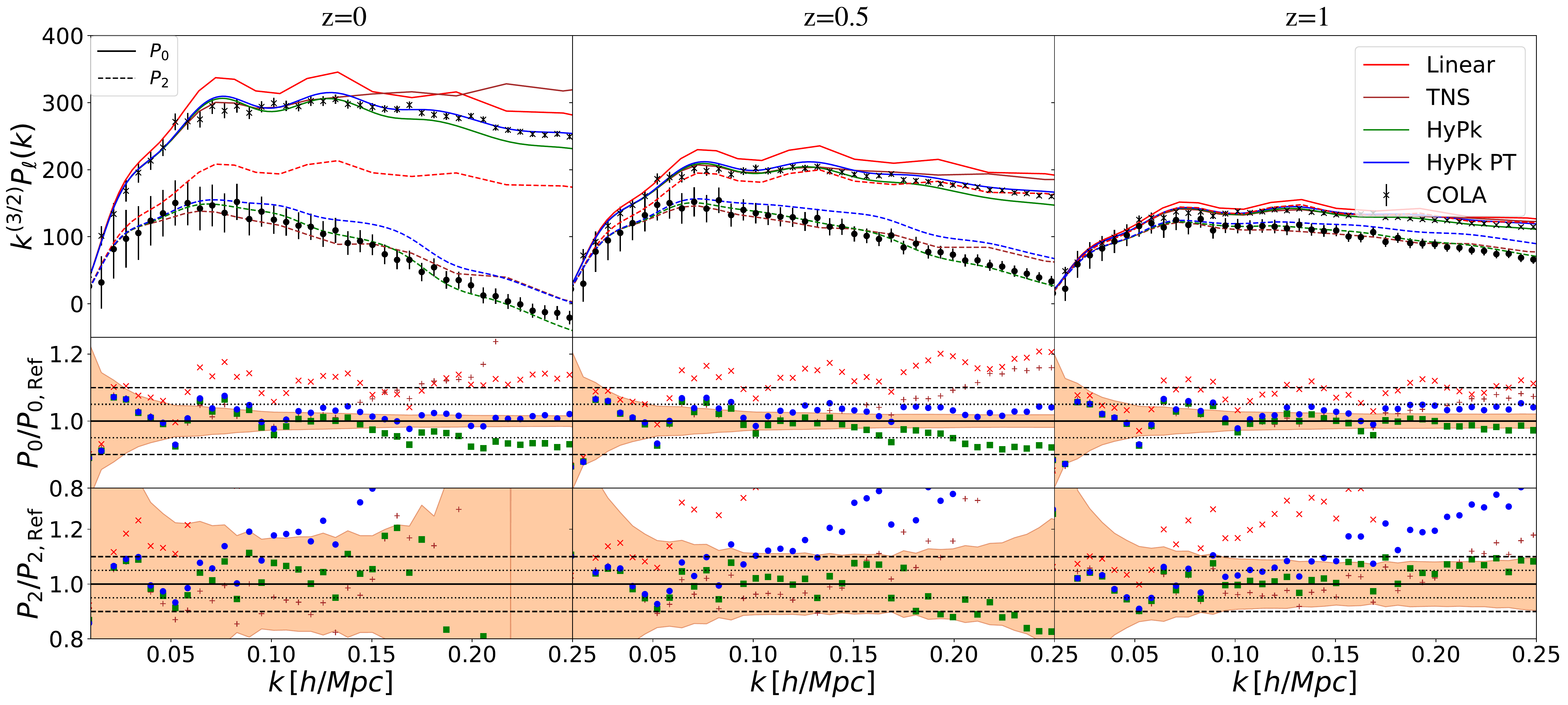}
  \caption[]{{\bf Redshift space multipoles in DGP}: Same as \figuretext~\ref{p02_grb} but for DGP gravity. For HyPk we use the reaction corrected halofit formula.}
\label{p02_dgpb}
\end{figure}

\begin{figure}[H]
\centering
  \includegraphics[width=\textwidth]{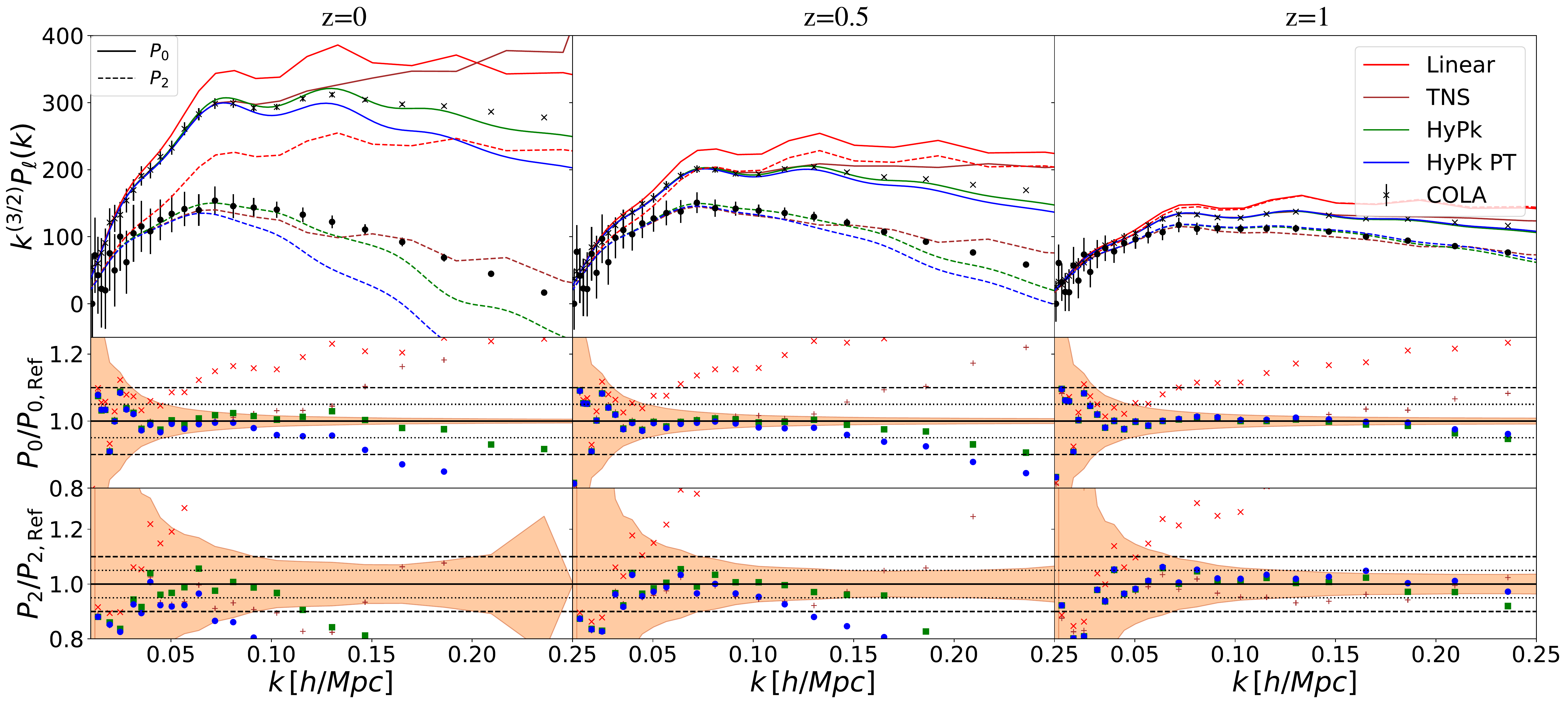}
  \caption[]{{\bf Redshift space multipoles in $f(R)$}: Same as \figuretext~\ref{p02_grb} but for $f(R)$ gravity. For HyPk we use the reaction corrected halofit formula.}
\label{p02_f4b}
\end{figure}

\begin{figure}[H]
\centering
  \includegraphics[width=\textwidth]{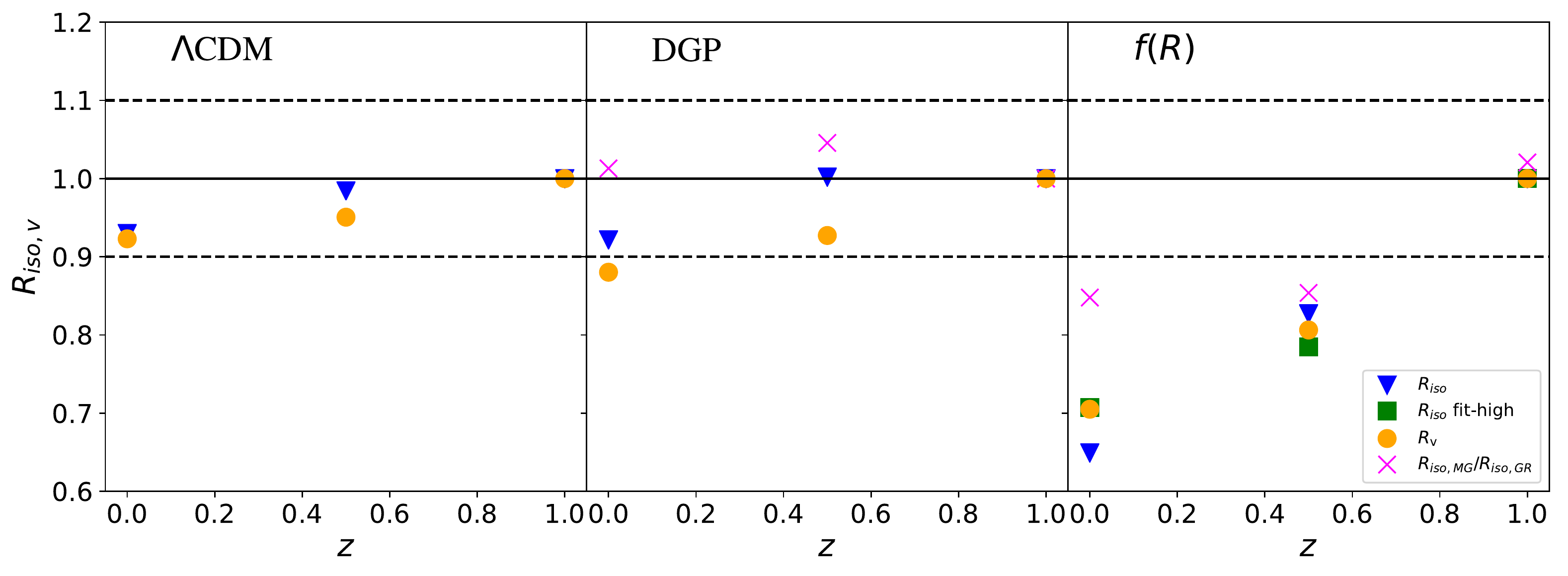}
  \caption[]{Ratio of fitted $\sigma_{\rm iso}$ and $\sigma_v$ to the predicted value from 1-loop perturbation theory $\sigma_{\rm iso,pt}$, normalised to 1 at $z=1$ (\eqtext~\ref{sigisov}). This ratio is plotted against redshift for $\Lambda$CDM, DGP and $f(R)$ from left to right. The TNS fit for $\sigma_v$ is shown as orange circles while the best fit $\sigma_{\rm iso}$ is shown as blue triangles. For $f(R)$ we also include the best fit at the higher $k_{\rm max}$ corresponding to $\chi^2_{\rm red}(k_{\rm max})\approx 1$ (green squares). We also show the ratio of $R_{\rm iso,v}$ in the modified gravity theories to the same quantity computed in GR as magenta crosses.}
\label{sigiso}
\end{figure}

Moving into redshift space, we compare the predictions of the hybrid approach against the COLA monopole, $P_0$, and quadrupole, $P_2$, and further compare to the well established TNS model \cite{Taruya:2010mx}. To fit the RSD free parameters $\sigma_{\rm iso}$ of HyPk and $\sigma_v$ of the TNS model, we perform a minimized $\chi_{\rm red}^2$ procedure on the COLA multipoles. We minimize the following function:
\begin{eqnarray}
\chi^2_{\rm red}(k_{\rm max}) & = \frac{1}{N_{\rm dof}}\sum\limits_{k=k_{\rm min}}^{k_{\rm max}} \sum\limits_{\ell,\ell'=0,2} \left[P^{S}_{\ell,{\rm data}}(k)-P^{S}_{\ell,{\rm model}}(k)\right]\nonumber \\ & \times \mbox{Cov}^{-1}_{\ell,\ell'}(k)\left[P^{S}_{\ell',{\rm data}}(k)-P^{S}_{\ell',{\rm model}}(k)\right],
\label{covarianceeqn}
\end{eqnarray}
where $\mbox{Cov}_{\ell,\ell'}$ is th Gaussian covariance matrix between the different multipoles, computed using linear theory, and $k_{\rm min}=0.009 \, h/{\rm Mpc}$.  We direct the reader to appendix C of \cite{Taruya:2010mx} for explicit expressions for $\mbox{Cov}_{\ell,\ell'}$. The number of degrees of freedom $N_{\rm dof}$ is given by $N_{\rm dof} = 2\times N_{\rm bins} - N_{\rm params}$, where $N_{\rm bins}$ is the number of $k-$bins used in the summation and $N_{\rm params}$ is the number of free parameters in the theoretical model. Here, $N_{\rm params} = 1$. {\B Again, we assume a survey volume of $V=4{\rm \, Gpc^3}/h^3$,  particle number density of $\bar{n}=10^{-3}h^3/{\rm Mpc}^3$, and linear bias $b_1=1$  in the covariance matrix.}
\\
\\
We begin by fitting the TNS model. To do this, we find the value of $\sigma_v$ for which $\chi^2_{\rm red}$ is minimal for a given $k_{\rm max}$. We then increase the value of $k_{\rm max}$ until $\chi^2_{\rm red,min}(k_{\rm max}) \geq 1$ as a good fit to the data is indicated when $\chi^2_{\rm red} \approx 1$. We adopt this $k_{\rm max}$ determined using the TNS model when fitting for $\sigma_{\rm iso}$ using the HyPk model; we then compare with TNS. We present a summary of the $k_{\rm max}$, the minimised $\chi^2_{\rm min}$ and values of the best fit parameters in \tabletext~\ref{fittable}. We also show the 1-loop prediction for $\sigma_{\rm iso}$, denoted as $\sigma_{\rm iso, pt}$. For GR and DGP we find similar $\chi^2$ values for TNS and HyPk at the TNS-chosen $k_{\rm max}$ but for $f(R)$ the hybrid model is able to fit the data significantly better than TNS, and hence we push it to a higher $k_{\rm max}$ as indicated in the $f(R)$ section of \tabletext~\ref{fittable}. 
\\
\\
\tabletext~\ref{fittable} highlights that given the current form of HyPk, the range of scales it is applicable at are comparable to the TNS model at all redshifts considered. The merits of the model lie in the use of the fully non-linear power spectrum as input; one might expect this to result in a model that is accurate to much smaller scales, but this depends on limiting factors coming from the treatment of the streaming model PDF and its ingredients. The  advantages of HyPk as well as various ways to improve its ability to reach smaller scales are highlighted in Sec.~\ref{sec:conclusion}.
\begin{table}[h]
\centering
\caption{{\bf Fitting information:} $k_{\rm max} [h/{\rm Mpc}]$, number of degrees of freedom $N_{\rm dof}$ and model parameters $\sigma_{v},\sigma_{\rm iso}$ and $\sigma_{\rm iso,pt}$ $[{\rm Mpc}/h]$ as well as the $\chi^2_{\rm min} (= N_{\rm dof} \chi^2_{\rm red}$) indicated in brackets. For $f(R)$ we also include the higher $k_{\rm max}$ and the respective $\sigma_{\rm iso}$ fits determined by the criterion of $\chi^2_{\rm red} \approx 1$. We note that, as mentioned in the main text, $\sigma_{\rm iso,pt}$ is not a  fitting parameter but a 1-loop prediction.}
\begin{tabular}{| c || c | c | c | c | c |}
 \multicolumn{1}{c}{{\bf $\Lambda$CDM}}  & \multicolumn{1}{c}{}   & \multicolumn{1}{c}{} & \multicolumn{1}{c}{{\bf TNS}} & \multicolumn{1}{c}{{\bf HyPk}} & \multicolumn{1}{c}{{\bf HyPk}}\\ \hline 
  {\bf z} &  $\boldsymbol{k}_{\rm max}$ & $\boldsymbol{N}_{\rm dof}$ & $\boldsymbol{\sigma}_v(\chi^2_{\rm min})$ & $\boldsymbol{\sigma}_{\rm iso} (\chi^2_{\rm min})$ & $\boldsymbol{\sigma}_{\rm iso, pt}(\chi^2)$ \\ \hline 
 $0$ & 0.132 & 41 & 6.8(47) & 4.6 (56) & 3.5 (100)  \\ \hline 
 $0.5$ & 0.150 & 47 &6.3(54) & 4.4 (56) & 3.1 (149)  \\ \hline 
 $1$ & 0.187 & 59 &4.9(65) & 3.3 (80) & 2.3 (182)  \\ \hline 
\multicolumn{6}{l}{{\bf DGP}} \\ \hline
 $0$ & 0.132 & 41 &8.0(46) & 5.4 (49) & 4.5 (84)  \\ \hline 
 $0.5$ & 0.138 & 43 &7.2(51) & 5.0 (44) & 3.9 (98)  \\ \hline 
 $1$ & 0.181 & 57 &5.5(60) & 3.5 (75) & 2.7 (147)  \\ \hline 
 \multicolumn{6}{l}{{\bf \emph f(R)}} \\ \hline
 $0$ & 0.115 & 45 &7.5(54) &  4.3 (28) & 6.3 (167)  \\ \hline 
 $0.5$ & 0.130& 47 &7.0(39) & 4.5 (21) & 5.2 (46)  \\ \hline 
 $1$ & 0.165  & 51 &5.9(60) & 3.7 (25) & 3.5 (31)  \\ \hline
  \multicolumn{6}{l}{{\bf \emph f(R)- higher $\boldsymbol{k}_{\rm max}$ fits}} \\ \hline
 $0$ & 0.130 & 47 &- &  4.4(52) & -  \\ \hline 
 $0.5$ &0.186 & 53 &- & 4.0(42) & - \\ \hline 
 $1$ & 0.209  & 55 &- & 3.5(53) & -  \\ \hline

\end{tabular}
\label{fittable}
\end{table}
\\
\\
We show the results for GR in \figuretext~\ref{p02_grb}. The upper panels show the monopole and quadrupole while the bottom panels show the ratios of the theoretical predictions with the COLA data. We immediately see that the monopole prediction for all models is very consistent with the COLA measurement, being $10\%$ accurate for $k\leq 0.25 \, h/{\rm Mpc}$. The quadrupole does worse, especially at lower redshift where it is more severely damped due to non-linear velocity dispersions. We also see that the hybrid model without any free parameters, HyPk PT,  underpredicts the velocity dispersion, indicated by a large quadrupole at quasi non-linear scales compared to the measurements. The results for DGP are qualitatively similar and are shown in \figuretext~\ref{p02_dgpb}. 
\\
\\
\figuretext~\ref{p02_f4b} shows the results for $f(R)$. In this case we clearly see that HyPk tends to do better than TNS, also reflected in the $\chi^2$ values shown in \tabletext~\ref{fittable}. Due to this behaviour, at the bottom of  \tabletext~\ref{fittable}, we also show a fit to a larger $k_{\rm max}$, found using the rough criterion of $\chi^2_{\rm red} \approx 1$. Note that \figuretext~\ref{p02_f4b} shows the spectra with the low $k_{\rm max}$ fits ($k_{\rm max}$ for HyPk and TNS are the same). {\B A key point here is the strong scale dependant growth factors of $f(R)$ and large velocity modifications \cite{Li:2012by,Jennings:2012pt}. In the monopole, scale dependencies will act to enhance power while strong velocity contributions will affect the damping induced by the fingers-of-god, mostly picked up by higher order multipoles. This causes the single damping parameter of the TNS model, $\sigma_v$, to be less effective in modelling both multipoles simultaneously. On the other  hand, HyPk contains the full non-linear density information which allows it to capture $f(R)$ scale dependencies more accurately. In doing so, the damping parameter, $\sigma_{\rm iso}$, can then better fit the higher order multipoles. This is consistent with \figuretext~\ref{p02_f4b}, where both HyPk multipoles (green) under-predict the simulations at high $k$. On the other hand, the TNS model over-predicts the monopole and under-predicts the quadrupole at intermediate scales with the best fit $\sigma_v$.}
\\
\\
Before proceeding, we make a few remarks. Although not included in the text, we have compared the RSD multipoles of the HyPk approach using an emulator and using the reaction for $P_{\rm \delta \delta}^{\rm NL}$ and find negligible difference between the two (for both $\Lambda$CDM and $f(R)$). Further, we have also considered the case where the linear power spectrum is used for $\xi$. We find this introduces significant inaccuracies into the RSD multipoles which cannot be effectively absorbed into the free parameter $\sigma_{\rm iso}$. These tests show that an accurate non-linear spectrum is key for this approach to work well. In addition, we have checked the fully non-linear ansatz for $v_{12}$ at $z=0$ \cite{Juszkiewicz:1998xf}. This showed minimal improvement in the quadrupole over the 1-loop result and even less improvement for the monopole. This indicates that most of the RSD information lies in $\sigma_{12}$ as well as the form of the PDF. We confirmed this by checking that going from the linear prescription of $\sigma_{12}$ to the 1-loop prescription significantly improves the quadrupole predictions. We will return to this point in {\B \sectext~\ref{sec:halos} and \sectext~\ref{sec:conclusion} }.  
\\
\\
Finally, we investigate the variation of the best fit values for $\sigma_{\rm iso}$ and $\sigma_v$ with redshift, comparing to the 1-loop PT prediction for $\sigma_{\rm iso}$. Such a comparison would be useful in the derivation of a redshift dependence and/or correction to the perturbative predictions for this parameter. This would be highly desirable in improving the model's ability to constrain cosmology or gravity. This comparison is shown in \figuretext~\ref{sigiso}. In particular, we show the ratio of the fit parameters with the 1-loop prediction $\sigma_{\rm iso,pt}$, normalised to 1 at $z=1$:
\begin{equation}
    R_{\rm iso, {\rm v}} = \left[\frac{\sigma_{\rm iso,{\rm v}}}{\sigma_{\rm iso,pt}}\right] \times \left[\frac{\sigma_{\rm iso,{\rm v}}}{\sigma_{\rm iso,pt}}\right]^{-1}_{z=1}.
   \label{sigisov}
\end{equation}
We note that for $\Lambda$CDM and DGP we have an approximate linear offset of the fitted values from the 1-loop prediction, with the 1-loop prediction underestimating the fitted values. For the $f(R)$ case, which has a scale dependent growth, this is no longer true, and we see larger deviations from the predictions and the fits at low redshift. Since in practice we are interested in $ z\geq 0.5$, this suggests one can possibly calibrate the fits at a given redshift for theories with scale independent growth. Further, we show the ratio $R_{\rm iso,v}$ in the modified gravity theories to the same quantity computed in GR as magenta crosses. Again for DGP, we see the ratio is almost constant further indicating that one can calibrate the fit for similar modifications to gravity. Note for $f(R)$, $R_{\rm  iso,{\rm v}}$ uses the high $k_{\rm max}$ fit of $\sigma_{\rm iso}$.

\subsection{Testing HyPk against {\it N}-body halos in $\Lambda$CDM} \label{sec:halos}
{\B
In this subsection we extend our comparisons by making use of a fully non-linear {\it N}-body simulation within $\Lambda$CDM. We also consider dark matter halos, which are biased tracers of the underlying dark matter distribution. This will provide a more defining test of the capabilities of HyPk within the context of a galaxy spectroscopic survey. 
\\
\\
For this analysis we relied on the BigMDPL simulation: a publicly available\footnote{https://www.cosmosim.org/cms/simulations/bigmdpl/} {\it N}-body simulation that followed the evolution of $N_{\rm part}=3840^3$ particles in a box of size $B=2500$ Mpc$/h$. The cosmological parameters used in this simulation is given by $h=0.678,\Omega_b = 0.0482,\Omega_M = 0.3071,n_s=0.96,\sigma_8 = 0.8228$. From the Rockstar halo catalogues of this simulation we selected all halos with mass $M > 10^{13}M_{\odot}/h$ at redshift $z=0.5$ and $z=1$ which corresponds to a co-moving number density of $\bar{n}(z=0.5) = 3.2\cdot 10^{-4} (h/\text{Mpc})^3$ and $\bar{n}(z=1) = 2.2\cdot 10^{-4} (h/\text{Mpc})^3$. The linear bias measured from these samples was found to be $b_1(z=0.5) = 2.04$ and $b_1(z=1) = 2.77$. The advantage of this simulation is that it has a very large volume, $15.625~\text{Gpc}^3/h^3$, and high spatial resolution allowing for accurate extraction of power-spectra and correlation functions. The downsides is that it just has one single realisation of the initial conditions so cosmic variance is still an issue for the largest pair separations and we also don’t have access to the dark matter particles. 
Additionally, going from dark matter to halos, there is additional stochastic noise which one needs to consider (see for example \cite{Beutler:2013yhm}). To account for this, the HyPk model will include a stochastic noise parameter, in exactly the same way as done in the TNS model. This is described next. 
\\
\\
We will consider three particular models for the redshift space power spectrum for biased tracers in this section. The first will be the simplest upgrade of the HyPk model used in section~\ref{sec:rsddm} to include linear bias (we refer the reader to \sectext~\ref{sec:bias} and to \cite{Reid:2011ar,Bose:2017dtl} for details on how this parameter, $b_1$, enters the modelling). Explicitly, this will be based on the GSM with $\xi(r)$ given by halofit and $v_{12}$ and $\sigma_{12}$ given by their 1-loop predictions. We fix the linear bias to that measured from the simulations. The isotropic contribution to the velocity dispersion $\sigma_{\rm iso}$ is fit to the power spectrum multipoles as done in the previous section. As mentioned, we also consider an additional stochastic parameter, $N$, within this model which is included as an additive term to the monopole, i.e. $P_0 = P_0^{S} + N$ where $P_0^{S}$ is given by equation \eqtext~\eqref{finalft}. This parameter is also fit to the multipoles. We will call this model for the power spectrum multipoles HyPk-Th and will represent the most theoretically general means of modelling the halo power spectrum multipoles with the hybrid approach. This model has 2 degrees of freedom, $N$ and $\sigma_{\rm iso}$.
\\
\\
The second model will be the TNS model used in the previous section but now equipped with the bias model of \cite{McDonald:2009dh} (see \eqtext~\eqref{redshiftps}), which gives the full model 4 nuisance parameters, 1 RSD parameter, $\sigma_v$, and 3 bias parameters $\{b_1,b_2,N\}$, where again $b_1$ is the linear bias, $b_2$ is second order bias, and $N$ is a stochastic/shot noise term. This model is very similar to the one used in the recent BOSS analysis \cite{Beutler:2016arn}, with some  differences as outlined in \cite{Markovic:2019sva,Bose:2019psj}. We will also fix $b_1$ to that measured from the simulations leaving this model with a total of 3 degrees of freedom. 
\\
\\
Finally, we will also compare HyPk with halo velocity statistics ($\sigma_{12}$ and $v_{12}$) and halo correlation function ($\xi_h$) as measured from the simulations themselves. This will represent a best-possible-case, where we have perfect models for all the GSM components. Performance of this model will largely reflect the validity of the Gaussian PDF ansatz used in equation~\eqref{eq:steamingmodel}, and the effects of Fourier transforming the Gaussian streaming model. We also include the same stochastic parameter, $N$, as in HyPk-Th, leaving this model with a single degree of freedom.
\\
\\
All free model parameters are fit to the power spectrum multipoles by minimising the likelihood given in equation~\eqref{covarianceeqn}. We use a Gaussian covariance matrix with $b_1$, $V$ and $\bar{n}$ as given by the BigMDPL simulation. We fit the nuisance parameters using scales up to $k_{\rm max}$, chosen such that the reduced $\chi^2$ is unity \footnote{For more details on the fitting procedure and choice of fitting range, we refer the reader to  \cite{Bose:2019ywu}.} as was done in the previous section. 
\\
\\
We begin by comparing the theoretical predictions for all GSM components with the {\it N}-body halo measurements. The first two moments of the pairwise velocity distribution are shown in \figuretext~\ref{hcompz05} and \figuretext~\ref{hcompz1} at $z=0.5$ and $z=1$ respectively. As noted in \cite{Reid:2011ar}, the 1-loop prediction for the mean pairwise infall velocity and perpendicular component of the pairwise infall velocity dispersion are very accurate, whereas the parallel component deviates at quasi non-linear scales. 
\begin{figure}[H]
\centering
  \includegraphics[width=\textwidth]{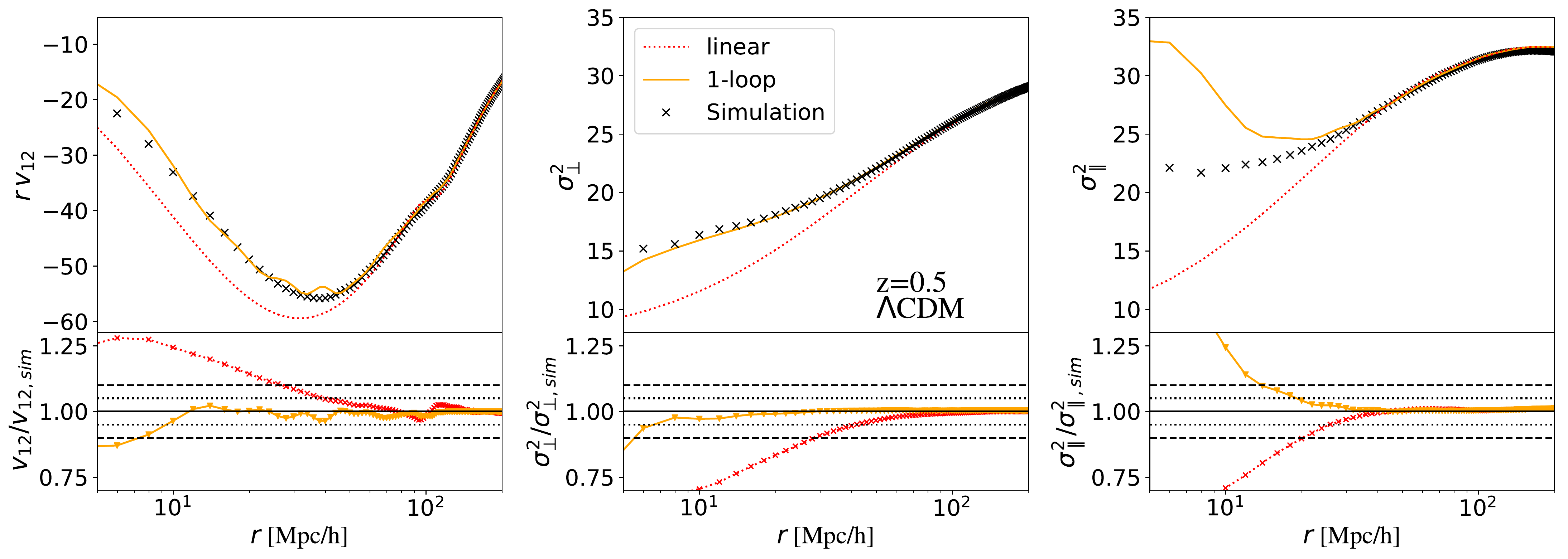}
  \caption[]{{\bf Gaussian PDF components in $\Lambda$CDM at $z=0.5$}: Top panels show the various prescriptions for the GSM components, i.e. linear (red dotted line) and 1-loop (orange solid line) against the {\it N}-body measurements (black crosses). The bottom panels show the ratio of the top panel predictions to the {\it N}-body measurements. The black dotted and dashed lines indicate $5\%$ and $10\%$ deviation, respectively. The left panel shows the mean pairwise infall velocity $v_{12}$ while the middle and left panels show the perpendicular and parallel components of the pairwise infall velocity dispersion. We have subtracted the mean infall velocity parallel to the line of sight from $\sigma_\parallel$ to get the dispersion about the mean. We have also shifted both the linear and 1-loop predictions for $\sigma_\parallel$ and $\sigma_\bot$ by a constant value to agree with the simulation measurements at large scales.} 
\label{hcompz05}
\end{figure}
\begin{figure}[H]
\centering
  \includegraphics[width=\textwidth]{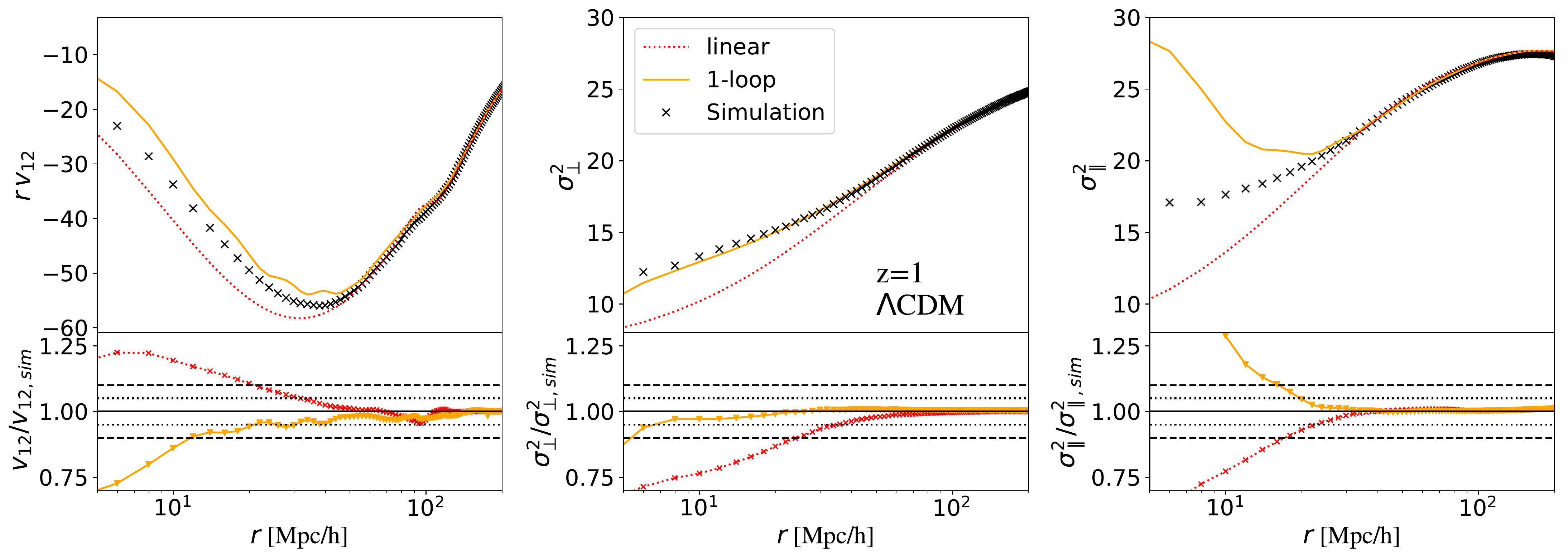}
  \caption[]{Same as \figuretext~\ref{hcompz05} but  at $z=1$.} 
\label{hcompz1}
\end{figure}
Next, \figuretext~\ref{xicompz05} and \figuretext~\ref{xicompz1} show the theoretical predictions for the halo power spectrum and its Fourier transform, the correlation function, at $z=0.5$ and $z=1$ respectively. We additionally show the Euclid emulator predictions here for reference. We find that the halofit and emulator equipped with linear bias are largely consistent in both Fourier and real space. In real space, these predictions remain within $10\%$ of the simulations for a very wide range of scales down to $r\sim 10{\rm Mpc}/h$, with the exception of the BAO scale. Here the emulator performs marginally better than halofit, but still shows up to $15\%$ deviations from the {\it N}-body  measurements.
\begin{figure}[H]
\centering
  \includegraphics[width=\textwidth]{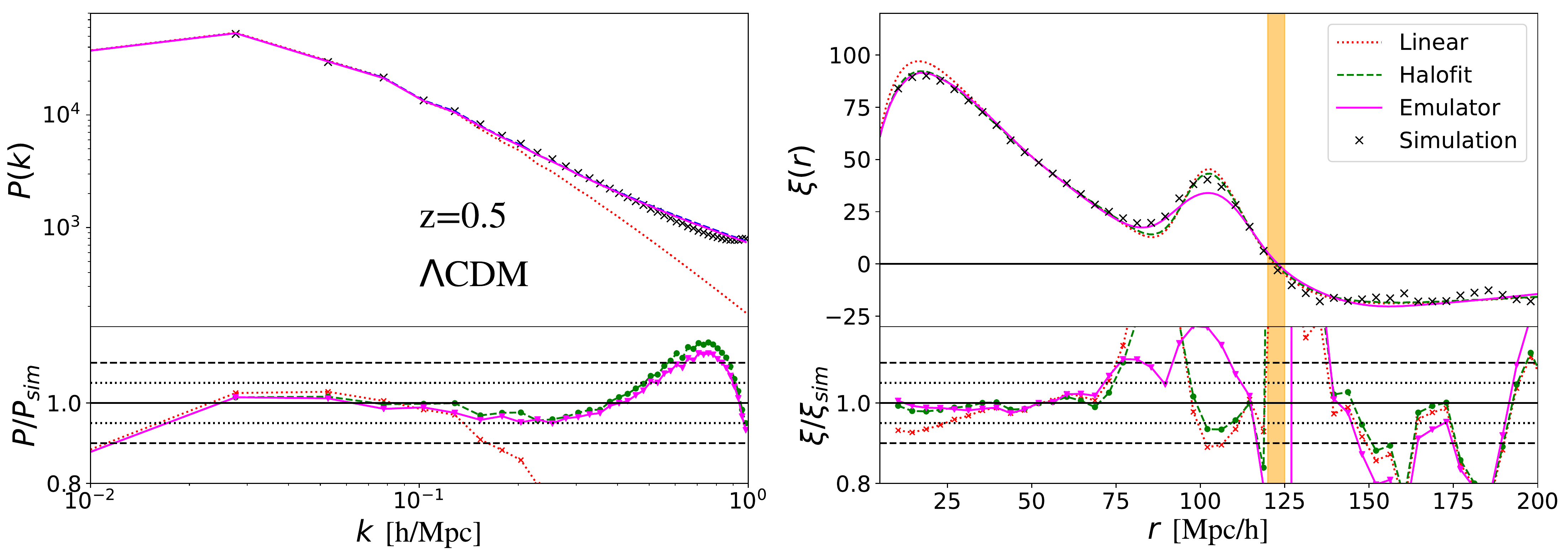}
  \caption[]{{\bf Halo power spectrum and correlation function  in $\Lambda$CDM at $z=0.5$}: Top panels show the various predictions for the 2-point halo clustering, i.e. linear (red dotted line), halofit formula (green dashed line) and Euclid emulator (magenta solid line) against the {\it N}-body measurements (black crosses). The bottom panels show the ratio of the top panel prescriptions to the {\it N}-body measurements. The black dotted and dashed lines indicate $5\%$ and $10\%$ deviation, respectively. The left panel shows the real space halo power spectrum while the right panel show the Fourier transform of the left panel predictions and the {\it N}-body measurement. The orange band in the right panel highlights a 0-crossing which results in large deviations in the ratio.} 
\label{xicompz05}
\end{figure}
\begin{figure}[H]
\centering
  \includegraphics[width=\textwidth]{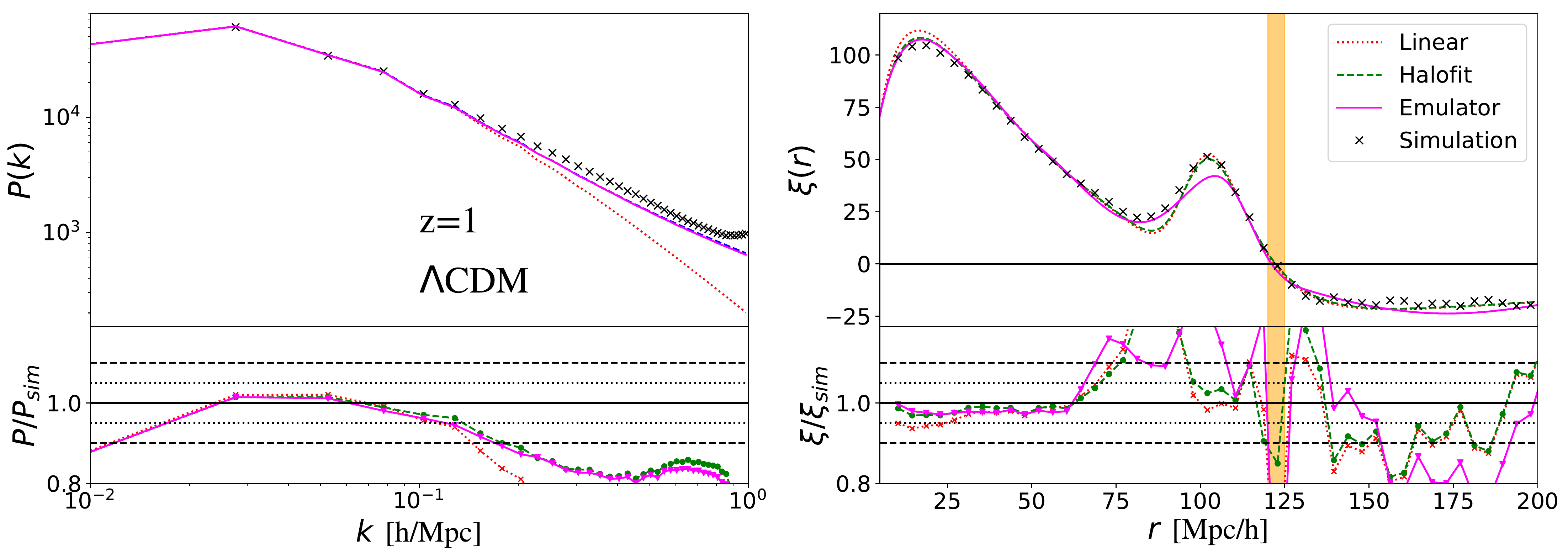}
  \caption[]{Same as \figuretext~\ref{xicompz05} but  at $z=1$.}
\label{xicompz1}
\end{figure}

We now move to redshift space. \figuretext~\ref{xi02comp_halos} shows the redshift space correlation function comparisons  at $z=0.5$ and $z=1$ in the range $10 \leq r \leq 200 ~{\rm Mpc}/h$. We find that the GSM model with theoretical predictions for $\xi$, $v_{12}$ and $\sigma_{12}$ matches the simulation measurements of the monopole to within $5\%$ at all scales except around the baryon acoustic oscillation (BAO) bump, where deviations become larger than $10\%$. This is similar to what was observed on the right hand side of \figuretext~\ref{xicompz05} and \figuretext~\ref{xicompz1}, a known inaccuracy of the halofit formula \cite{Smith2003,Takahashi:2012em}. It also does very well at modelling the quadrupole down to $r\sim 30{\rm Mpc}/h$ at $z=0.5$ and slightly further at $z=1$. The GSM model with simulation measurements does very well  at modelling both multipoles at all scales including the BAO scale except the quadrupole at large separations, where statistical errors in the measurements are expected to be large.
\\
\\
Next, we move to Fourier space. \figuretext~\ref{p02comp_halos} shows the Fourier transforms of all curves in \figuretext~\ref{xi02comp_halos} as well as the best fit TNS model. Further, we re-fit $\sigma_{\rm iso}$ to best fit the spectra so as to more directly compare with TNS. We find that TNS and HyPk-Th do equally well in modelling the monopole (quadrupole), keeping within a $3 \, (10) \%$ percent of the measurements down to $k\approx 0.3h/{\rm Mpc}$ at both $z=0.5$ and $z=1$. HyPk-Sim does equally well in modelling the monopole and is more accurate that both HyPk-Th and TNS in modelling the quadrupole in $k\leq0.2 h/{\rm Mpc}$, but deviates at smaller scales. It should be noted that TNS benefits from an additional degree of freedom, $b_2$, when fitting the data. 
\\
\\
In \figuretext~\ref{p02comp_halos} we also show the monopole of HyPk-Sim with $N=0$. The large deviation from the measured monopole shows the stochastic term is essential in capturing it accurately. We have checked that this contribution is not a fault of the measurements by making a second measurement of a separate halo catalogue with twice the number of halos, which still results in a large failure of HyPk-Sim in modelling the monopole. The best fits for HyPk-Sim are found to be $N(z=0.5)=781$ and $N(z=1)=1270$. The best fits for HyPk-Th are consistent, being $N(z=0.5)=578$ and $N(z=1)=1150$. In comparison, the TNS model with $b_2=0$, fit up to the same $k_{\rm max}$, yields $N(z=0.5)=480$ and $N(z=1)=850$ which corresponds to $\sim 15\%$ of the shot-noise that we have subtracted from the simulation measurements. 
\\
\\
Finally, in \figuretext~\ref{chi2plot} we show the reduced $\chi^2$ statistic given by \eqtext~\eqref{covarianceeqn} for linear theory (Kaiser) and best fit HyPk-Th, TNS, HyPk-Sim \footnote{Recall that the models are fit at $k_{\rm max}$ where $\chi^2_{\rm red} \approx 1$.} as a function of the maximum wave number used in calculating the statistic. A $\chi_{\rm red}^2\approx 1$ is generally indicative of a good fit to the data without over-fitting. Again, HyPk and TNS do extremely well at modelling the halo multipoles, having $\chi^2_{\rm red} \leq 1$ up to $k_{\rm max} \approx 0.275h/{\rm Mpc}$ for both $z=0.5$ and $z=1$. For TNS, similar results were found in \cite{Bose:2019psj,Bose:2019ywu}. These results suggest a simple bias modelling within HyPk may offer a very good means of modelling the halo power spectrum multipoles.  
\\
\\
This being said, improvements can be made to HyPk's accuracy.  Modelling the infall velocity PDF down to very small scales, $r\leq 10{\rm \, Mpc}/h$, may be very important when taking the Fourier transform. Indeed, we find that one needs to integrate down to $s\approx 1 {\rm \, Mpc}/h$ in \eqtext~\eqref{finalft} to get reasonable results. On this point, we test the Gaussian PDF ansatz in Appendix~\ref{app:pdf} and find that Gaussianity becomes a poor approximation for $r\leq  10{\rm \, Mpc}/h$ \footnote{Using the full measured PDF in the streaming model and subsequent Fourier transform is both numerically challenging and computationally expensive, and so we make no such comparisons of the power spectrum multipoles here.}.
\\
\\
Further, for scales below $r\approx 10 {\rm \, Mpc}/h$ we do not have sufficient resolution to accurately measure $\xi_h(r)$ from the simulations and so extrapolate using the emulator with linear bias, which may cause some of the discrepancy we see between HyPk-Sim and the measurement of $P_0$. Such dependency on these non-linear scales is a clear disadvantage of the current HyPk approach. Very accurate predictions for these scales or using a phenomenological parametrisation are the two obvious ways of resolving this issue. Currently, a simple shot noise degree of freedom seems to greatly improve HyPk's accuracy. We discuss these points further in the conclusion. 
}
\begin{figure}[H]
\centering
  \includegraphics[width=\textwidth]{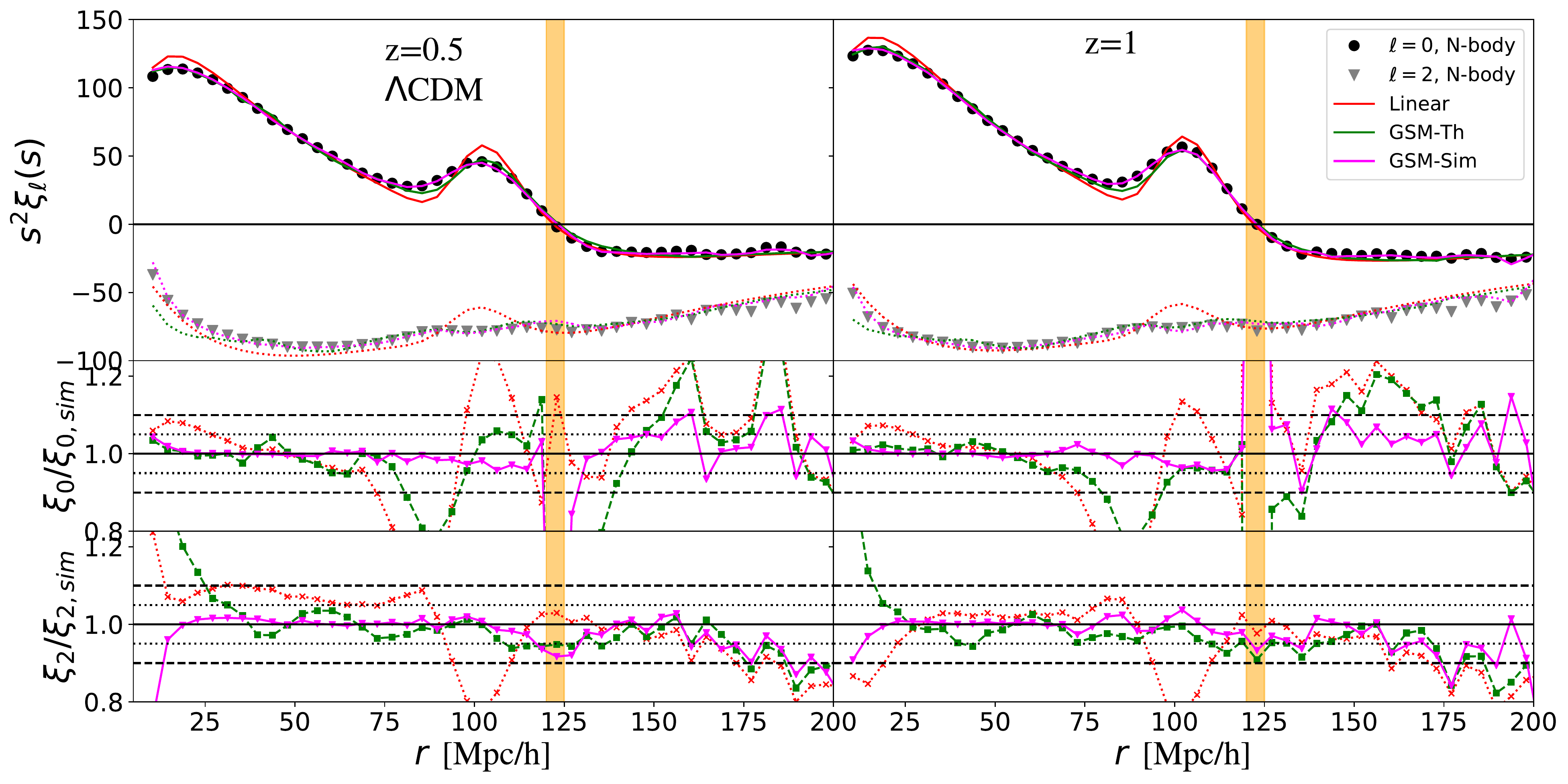}
  \caption[]{{\bf Redshift space correlation function multipoles in $\Lambda$CDM}: Top panels show the {\it N}-body monopole (black circles) and quadrupole (grey triangles) against the theoretical predictions. We show the theoretical predictions for the monopole (solid) and quadrupole (dotted) spectra using Kaiser (red), GSM with halofit $\xi(r)$ and 1-loop $v_{12}$ and $\sigma_{12}$ (green) and GSM with the simulation measurements for all GSM components (magenta). Middle and bottom panels show the ratio of the theoretical predictions to the {\it N}-body measurements for $\xi_0$ and $\xi_2$ respectively, with the black dotted and dashed lines indicating $5\%$ and $10\%$ deviation, respectively. The vertical orange band highlights a 0-crossing which results in large deviations in the ratio. For the GSM-Th predictions we set $\sigma_{\rm iso}$ to the large scale velocity dispersion measured from the simulations.} 
\label{xi02comp_halos}
\end{figure}
\begin{figure}[H]
\centering
  \includegraphics[width=\textwidth]{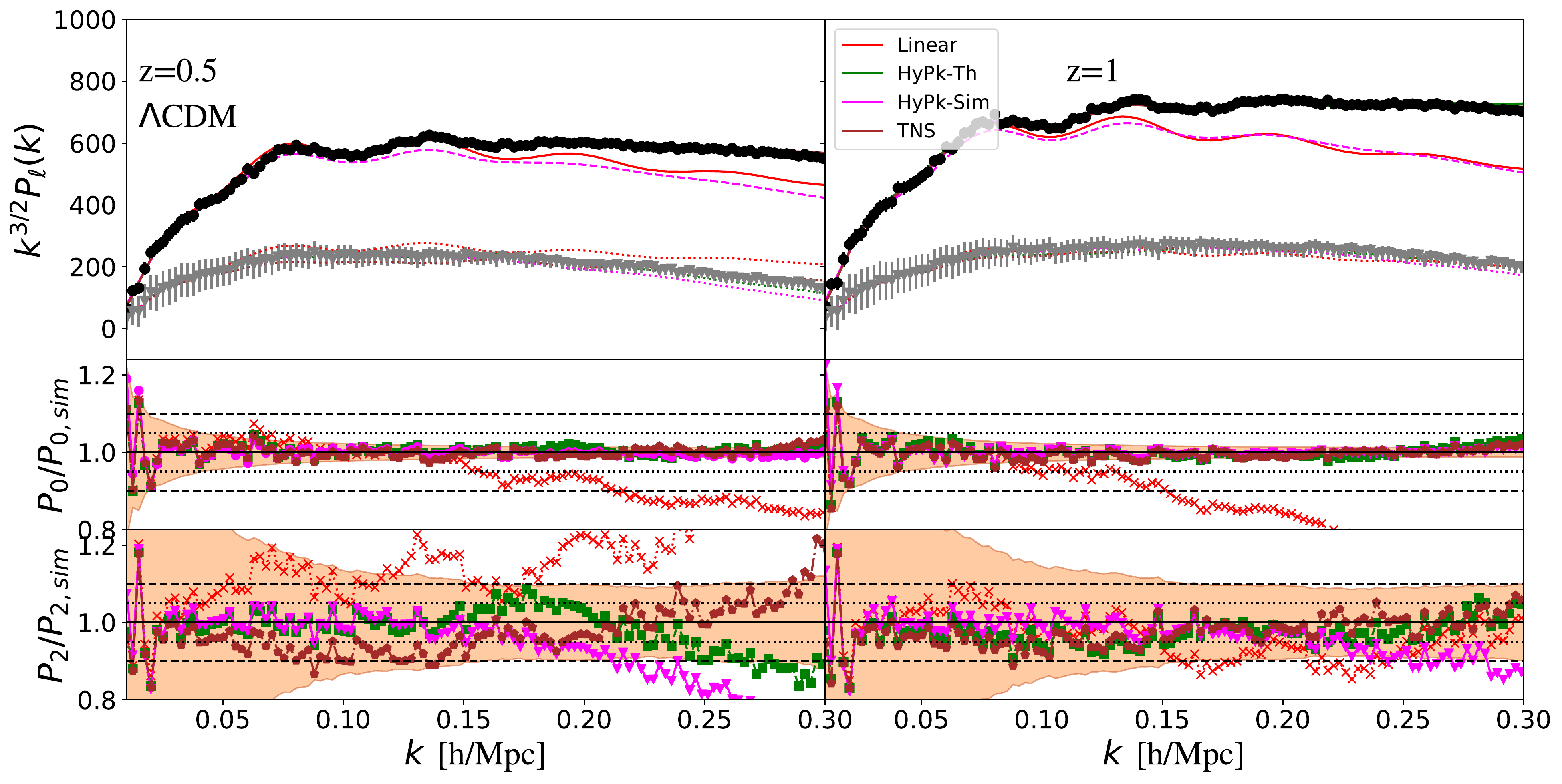}
  \caption[]{{\bf Redshift space power spectrum multipoles in $\Lambda$CDM}: Top panels show the {\it N}-body monopole (black circles) and quadrupole (grey triangles) against the theoretical predictions. We show the theoretical predictions for the monopole (solid) and quadrupole (dotted) spectra using Kaiser (red), HyPk with halofit $\xi(r)$ and 1-loop $v_{12}$ and $\sigma_{12}$ (green) and HyPk with the simulation measurements for all GSM components (magenta). The monopole prediction of HyPk-Sim without $N$ is shown as a magenta dashed curve.  Middle and bottom panels show the ratio of the theoretical predictions to the {\it N}-body measurements for $P_0$ and $P_2$ respectively, with the black dotted and dashed lines indicating $5\%$ and $10\%$ deviation, respectively. The error bars and bands are computed using twice the diagonal components of the linear theory covariance matrix used in \eqtext~\ref{covarianceeqn} ($2\sigma$-errors), with $b_1$, $V$ and $\bar{n}$ all given by the simulation specifications described in the main text.}
\label{p02comp_halos}
\end{figure}

\begin{figure}[H]
\centering
  \includegraphics[width=\textwidth]{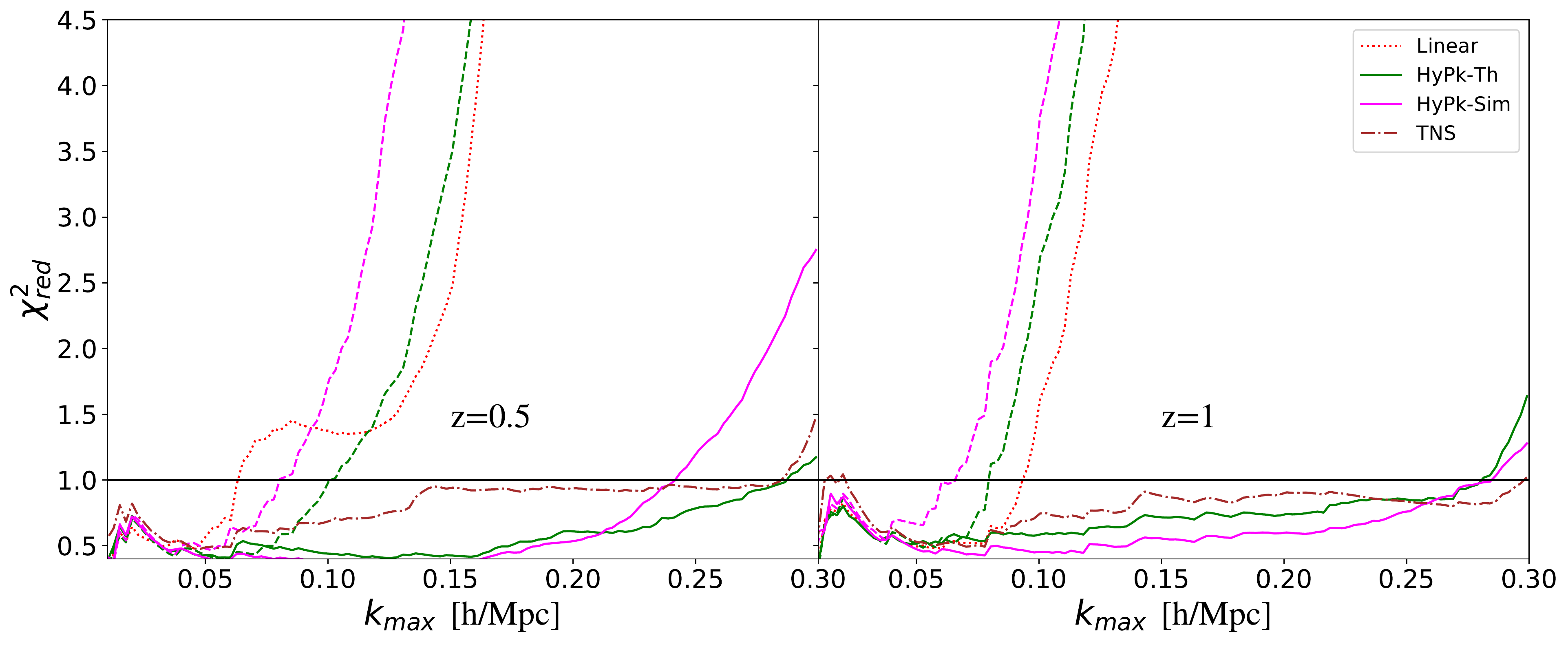}
  \caption[]{Reduced $\chi^2$ as a function of $k_{\rm max}$ for the models considered in \figuretext~\ref{p02comp_halos}. HyPk with and without shot noise term $N$ are shown as solid and dashed lines respectively.}
\label{chi2plot}
\end{figure}

\section{Conclusions}\label{sec:conclusion}
 In this paper we have presented a new, general approach to modeling the non-linear power spectrum in redshift space. We have built upon previous work \cite{Cataneo:2018cic} by taking the generalised non-linear, real space matter spectrum to redshift space, and also extended \cite{Bose:2017dtl}, taking the generalised perturbative prescription for the Gaussian streaming model back to Fourier space. 
\\
\\
For the GSM ingredients we focus on using 1-loop SPT for the Gaussian PDF and the FT of a fully non-linear matter power spectrum prediction coming from a joint halo model-SPT approach described in \cite{Cataneo:2018cic}. 
\\
\\
The {\bf key results} of this work are summarised below:
\begin{itemize}
    \item 
    The  Hybrid $P_\ell(k)$ (HyPk) approach is competitive with the TNS model in modelling the dark matter power spectrum monopole and quadrupole from COLA measurements for $\Lambda$CDM, DGP and $f(R)$ gravity at $z=0,0.5$ and $z=1$.
    \item
    The hybrid approach does significantly better than TNS when one considers $f(R)$ gravity. 
    \item
    Our results suggest that for some gravitational theories, including $\Lambda$CDM and DGP, the redshift dependence of $\sigma_{\rm iso}$ can be modelled using perturbation theory. This is not the case for theories with scale dependent growth like $f(R)$.
    \item
    {\B Modelling tracer bias within this hybrid approach is necessary and non-trivial. Promisingly, by comparing to a full {\it N}-body halo catalog in $\Lambda$CDM at $z=0.5$ and $z=1$, we find that including linear bias and shot noise parameter within the hybrid approach does competitively with TNS equipped with a McDonald $\&$ Roy bias model \cite{McDonald:2009dh}, despite the TNS having an additional free parameter, $b_2$.} 
    \item
    {\B The HyPk approach with a shot noise degree of freedom, similar to what is done in the TNS model, allows HyPk to predict the first two halo multipoles accurately down to $k \approx 0.275h/{\rm Mpc}$ at $z=0.5$ and $z=1$.}
    \item
    {\B The HyPk approach is sensitive to the very non-linear regime  due to the Fourier transform. Modelling the very non-linear regime and/or parametrising the physics on these scales will be crucial in improving the accuracy of this  approach.}
    \item
    {\B Currently, we do not find a need to go beyond the Gaussian approximation in the streaming model as HyPk performs equally well with current state of the art models for the redshift space power spectra.}
\end{itemize}
The key {\bf advantages} of the approach provided here are summarised below:
\begin{itemize}
    \item 
    The hybrid approach allows unification of clustering and lensing analyses by requiring a single prescription of the fully non-linear matter power spectrum in a consistent way  (same input matter power spectrum). As with the perturbative approach, it also allows one to be agnostic about gravity and dark energy modelling.  
    \item
     For dark matter it is competitive with the current state of the art RSD spectra model (TNS).
    \item
    Since it utilises the fully non-linear power spectrum, it is expected to do much better than current perturbative models for $\mu \rightarrow 0$. This will be of particular importance when looking at the angular power spectra \cite{Jalilvand:2019brk}.
    \item
    The approach also provides very direct means of improvement, to be discussed next. 
\end{itemize}
 {\B The main {\bf disadvantages} are the obvious dependency of the predictions on the redshift space correlation function over a very wide range of scales, including highly non-linear scales, as well as including tracer bias in a fully non-linear prescription for the real space, dark matter, correlation function $\xi(r)$. Both are unclear and necessary for this approach to be considered in any sort of data analysis.} Despite this, the range of scales that our proposed HyPk model is applicable at, seems comparable to the 1-loop SPT based TNS model {\B for dark matter and halos}. {\B We await a more in depth, statistical analysis to make strong claims.} Further, whereas it is unclear how to improve non-linear RSD modelling using perturbative techniques (see \cite{Blas:2013aba,Carlson:2009it,Konstandin:2019bay} for various limitations), the hybrid approach can be {\bf improved} in many ways. We list these below:
\begin{itemize}
     \item
     One can improve the prescriptions for $\sigma_{12}$ and $v_{12}$, the mean and standard deviation of the Gaussian PDF. Such non-linear prescriptions have been proposed in the literature \cite{Juszkiewicz:1998xf,Caldwell:1999uf,Falco:2012ud,Falco:2013zya,Lombriser:2012nn,Vlah:2016bcl,Valogiannis:2019nfz} and will be the focus of upcoming work. Particularly, $\sigma_{12}$ contains valuable RSD anisotropy information and hence a non-linear prescription for this quantity should be tested within the GSM framework. 
     \item
     One can implement a halo-model prescription for $\sigma_{12}$ and $v_{12}$ that would unify the framework under the halo-model, as the reaction for $P_{\delta \delta}^{\rm NL}$ is already based on this. This is one of our next key objectives. 
     \item
     Related to the previous point, one can include bias consistently using the halo-model, through halo-bias. This would give a fully predictable theoretical model. One could then parametrise inaccuracies using the simulation fits within the halo-model ingredients for example (mass function, virial concentration, and halo profile). 
     \item 
     One can go beyond the Gaussian PDF. This has been shown to greatly improve the streaming model predictions at small scales \cite{Bianchi:2014kba,Uhlemann:2015hqa,Bianchi:2016qen,Vlah:2016bcl}.
\end{itemize}
{\B In particular, in \cite{Valogiannis:2019nfz,Valogiannis:2019xed} it was shown that the GSM with a local Lagrangian bias scheme and a resummed-Lagrangian perturbation theory modelling for the real space correlation function can do very well in reproducing simulations results for $\xi_{0,2,4}$ within modified gravity. Using such a scheme for the velocity PDF components is left to a future work.}
\\
\\
We note that some proposals also aim to combine perturbative and halo-model approaches for the RSD spectra (e.g.  \cite{Vlah:2013lia,Okumura:2015fga}), but these typically introduce additional degrees of freedom to model more detailed redshift space anisotropies. Still, one can ask how well the model described in \cite{Okumura:2015fga} can do when combined with the reaction for the non-linear power spectrum - we aim to investigate this in future work. This method would also not need the double Fourier transform which can be sensitive to the non-linear regime.
\\
\\
In closing, we make some general remarks regarding the numerical tools used in this work; they are summarised in \tabletext~\ref{codes}. We have utilised all the codes listed for the results presented in this paper. For the comparison with simulations we use {\tt MG-PICOLA} and the rest of the codes are all involved in the {\tt HyPk} computation, depending on the model. We have written a {\tt Python} script\footnote{The code is available at \url{https://github.com/nebblu/HyPk}.} that combines outputs from {\tt MG-Copter} and {\tt ReACT}, and then applies the {\tt FFTLog} algorithm. This turns out to be very cumbersome and not suited for model fitting or parameter estimation. Ideally we would have all of these packages working under the same script. Further improvement of the HyPk approach and a development of a fast code for parameter estimation is a current goal of the authors.

\begin{table}[h]
\centering
\caption{Codes used in this work and their function. NL stands for non-linear.}
\begin{tabular}{| c || c | c | c | c | c |} \hline 
  Code &$P_{\delta \delta}$ & $P^S_{0,2}$  & $\sigma_{12}$, $v_{12}$ & $\mathcal{R}$ & FFT   \\ \hline \hline
  {\tt MG-PICOLA}\cite{Winther:2017jof} & NL & NL & - & - & -   \\ \hline
  {\tt EuclidEmulator} \cite{Knabenhans:2018cng} & NL** & - & - & - & -   \\ \hline 
  {\tt fRemulator}  \cite{Winther:2019mus} & NL*** & - & - & - & -   \\ \hline 
  {\tt MG-Copter}*\cite{Bose:2016qun,Bose:2017dtl} & halofit** & TNS & linear/1-loop & - & -   \\ \hline 
  {\tt ReACT}*\cite{react} & halofit** & - & - & \checkmark & -   \\ \hline 
  {\tt FFTLog}*\cite{Hamilton:1999uv}  & - & - & - & -&  \checkmark  \\ \hline 
  \multicolumn{6}{l}{\footnotesize   $*$ used in HyPk computation} \\
  \multicolumn{6}{l}{{\footnotesize   $**$   GR only}} \\
  \multicolumn{6}{l}{{\footnotesize  $***$  $f(R)$ only}}\\
  \end{tabular}
\label{codes}
\end{table}

%%%%%%%%%%%%%%%% %%%%%%%% %%%%%%%% 
\section*{Acknowledgments}
 BB and LL acknowledge support from the Swiss National Science Foundation (SNSF) Professorship grant No.~170547. AP is a UK Research and Innovation Future Leaders Fellow, grant MR/S016066/1, and also acknowledges support from the UK Science \& Technology Facilities Council through grant ST/S000437/1.
 SC acknowledges support from CNRS and CNES grants. QX and MC acknowledge support from the European Research Council under grant number 647112. SC and LL are grateful for the hospitality of the Institut Pascal during the final stages of this work and the support by ``P2IO LabEx (ANR-10-LABX-0038)'' in the framework ``Investissements d'Avenir'' (ANR-11-IDEX-0003-01) managed by the Agence Nationale de la Recherche (ANR, France). We acknowledge use of open source software \citep{scipy:2001,Hunter:2007,  mckinney-proc-scipy-2010, numpy:2011, Lewis:1999bs, Lesgourgues:2011re}. The CosmoSim database used in this paper is a service by the Leibniz-Institute for Astrophysics Potsdam (AIP).
The MultiDark database was developed in cooperation with the Spanish MultiDark Consolider Project CSD2009-00064. The authors would like to thank the anonymous referee for their very useful contributions.

\appendix \label{app:pdf}
{\B
\section{Halo pairwise velocity PDF comparisons in $\Lambda$CDM}
In this section we provide comparisons of the measured halo pairwise velocity probability distribution function from the BigMDPL Multidark {\it N}-body simulation \cite{2016MNRAS.457.4340K}  and the Gaussian ansatz used in \eqtext~\ref{eq:steamingmodel}. 
\\
\\
\figuretext~\ref{pdfz05} and \figuretext~\ref{pdfz1} contain comparisons of the measured pairwise velocity PDF with the Gaussian ansatz with various theoretical predictions for $v_{12}$ and $\sigma_{12}$ as well as the measured $v_{12}$ and $\sigma_{12}$. For the 1-loop prediction, we tune the isotropic velocity dispersion, $\sigma_{\rm iso}$, to best fit the simulation's velocity dispersion at large scales. The Gaussian ansatz does surprisingly well except at very non-linear scales, doing  worse for angles along the line of sight where the RSD effects are greatest. At these scales, we can see that the skewness becomes important, and to a lesser degree the higher order moments, a result consistent with \cite{Bianchi:2016qen}. It is also clear that the 1-loop prediction does very well in matching the simulation predictions within the Gaussian ansatz for angular scales that are not too close to the line of sight.
\begin{figure}[H]
\centering
  \includegraphics[width=\textwidth]{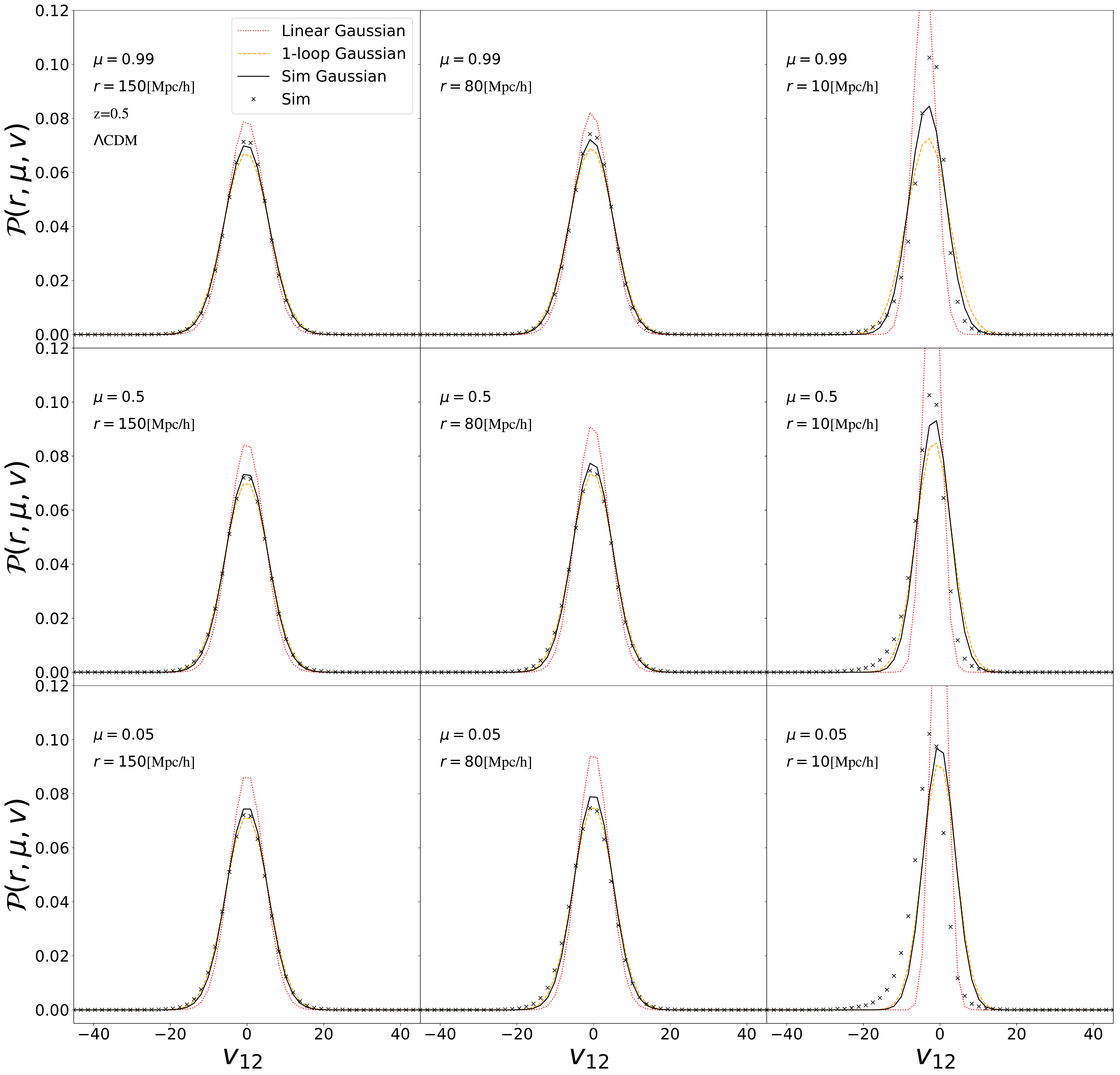}
  \caption[]{{\bf Pairwise velocity PDF in $\Lambda$CDM at $z=0.5$}: The pairwise velocity probability distribution function as measured from the {\it N}-body simulation (black crosses) compared to a Gaussian distribution with  mean and standard deviation as predicted by  linear theory (red dotted line), 1-loop perturbation theory (orange dashed line) and as measured from the simulations (black solid line). We show the 2D predictions for various values of angular and physical pair separation, $\mu$ and $r$. } 
\label{pdfz05}
\end{figure}
\begin{figure}[H]
\centering
  \includegraphics[width=\textwidth]{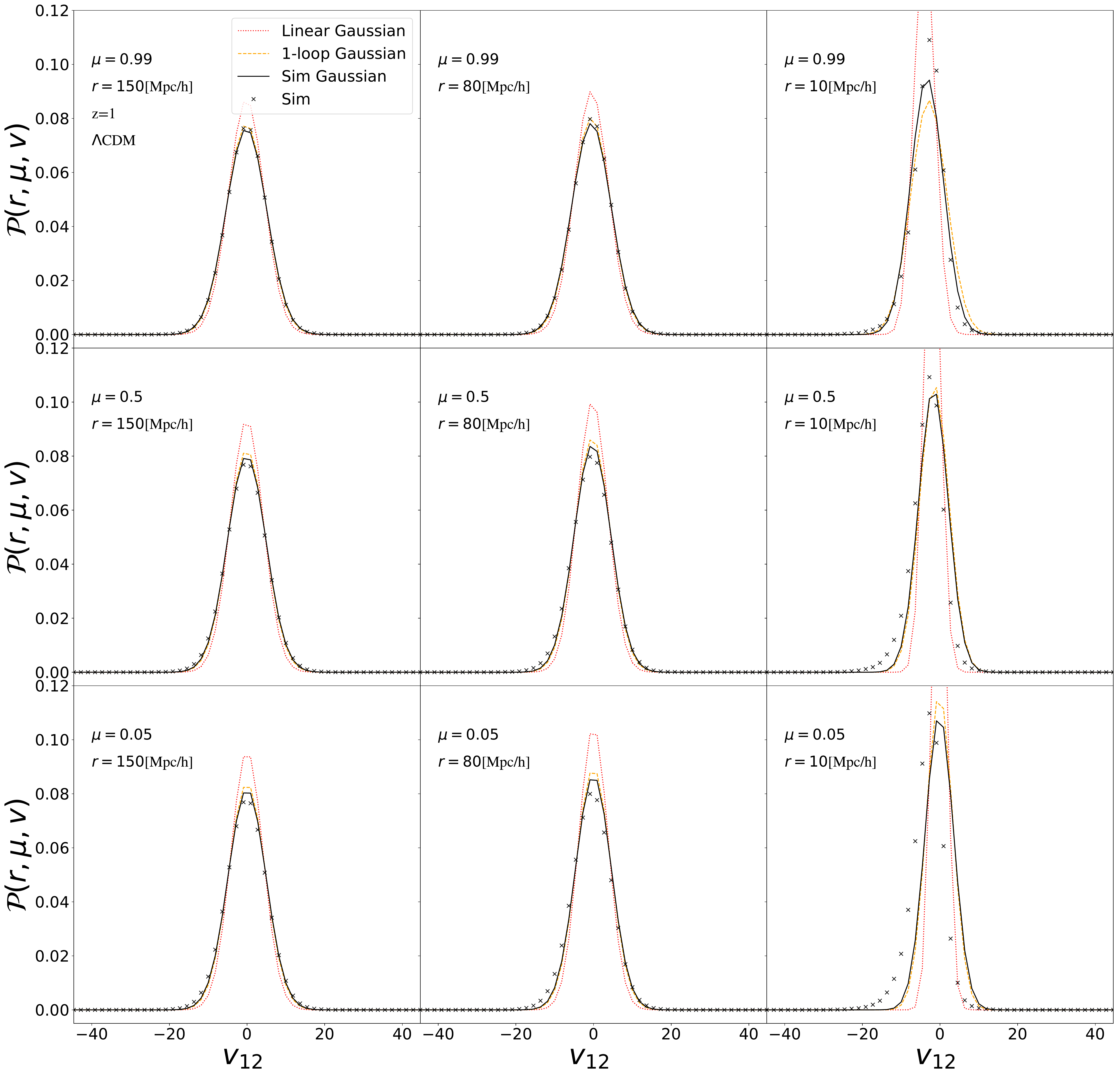}
  \caption[]{Same as \figuretext~\ref{pdfz05} but  at $z=1$.} 
\label{pdfz1}
\end{figure}

}
\bibliography{mybib} 
\bibliographystyle{ieeetr}

\end{document}